\documentstyle[12pt]{article}
\def\del        {  \partial  }
\def\half       {  {1\over 2}  }

\def\defint#1#2 {  \int_{#1}^{#2}  }
\def\rootof#1   {  \left( #1 \right)^{1/2}  }
\def\deldel#1   {  {\partial\over \partial #1}  }

\def\abs#1      {  \vert #1 \vert  }
\def\ie         {  {\it i.e.}      }
\def\evalat#1   {  \left\vert_{#1} \right. }
\def\re         { {\rm e}  }
\def\comma      {\, ,}
\def\period     {\, .}
\def\lsim    {\lower .65ex \hbox{\ $\stackrel{<}{\sim}$\ } }
\def\gsim    {\lower .65ex \hbox{\ $\stackrel{>}{\sim}$\ } }
\def\calO       { {\cal O} }

\def\calU       { {\cal U} }

\def\Ltilde{\tilde{L}}
\def\mndelta{\delta_{m+n,0}}
\def\calx{{\cal X}}
\def\oneoveri{{1 \over i}}
\def\overrttwo{{1\over \sqrt{2}}}
\def\Lzerotot{L^{tot}_0}
\def\Phizero{\Phi_0}
\def\overi{{1\over i}}
\def\xplus{{x^+}}
\def\xminus{{x^-}}
\def\yplus{{y^+}}

\def\delplus{\partial_+}
\def\delminus{\partial_-}
\def\rttwo{\sqrt{2}}
\def\Qbar{\bar{Q}}
\def\Pbar{\bar{P}}
\def\rphi{\varphi}% round phi
\def\thsig{\theta(\sigma)}
\def\thrho{\theta(\rho)}

\def\calUinv{\calU^{-1}}
\def\ft#1#2{{\textstyle{{#1}\over{#2}}}}
\def\a{\alpha}
\def\b{\beta}
\def\g{\gamma}\def\G{\Gamma}
\def\d{\delta}\def\D{\Delta}
\def\e{\epsilon}

\def\t{\theta}
\def\l{\lambda}
\def\m{\mu}

\def\n{\nu}
\def\p{\psi}
\def\r{\rho}
\def\s{\sigma}
\def\ta{\tau}

\def\ve{\varphi}

\def\xip{\xi^+} \def\xim{\xi^-}
\def\gp{{{\gamma^2}\over {8\pi}}}
\def\papertitlepage{\baselineskip 3ex \thispagestyle{empty}}
\def\Title#1{\vspace{1cm}\begin{center}
 {\Large\bf #1} \end{center}
\vspace{1cm}}
\def\Authors#1{\begin{center} {\it #1} \end{center}}
\def\Abstract{\vspace{1cm}\begin{center} {\large\bf Abstract}
           \end{center} \parmedskip}

\renewcommand{\thefootnote}{\fnsymbol{footnote}}
\renewenvironment{thebibliography}{\pagebreak[3]\par\vspace{0.6em}
\begin{flushleft}{\large \bf References}\end{flushleft}
\vspace{-1.0em}

\begin{enumerate}\if@twocolumn\baselineskip=0.6em\itemsep -0.2em
\else\itemsep -0.2em\fi\labelsep 0.1em}{\end{enumerate}}
\def\eqabegin         {  \begin{eqnarray}  }
\def\eqaend           {  \end{eqnarray}  }
\def\nn               {  \nonumber  }
\newcommand{\Subsection}[1]{\setcounter{equation}{0}%
                              \subsection{#1}}
\renewcommand{\theequation}{\arabic{section}.\arabic{subsection}.%
                            \arabic{equation}}
\def\bracetwo#1#2     {  \left\{ \begin{array}{l} #1 \\ #2 \end{array}
                         \right.  }
\def\bracetwocases#1#2#3#4  {   \left\{ \begin{array}{ll} #1 & \qquad #2 \\
                                 #3 & \qquad #4 \end{array} \right.  }
\def\bracebegin#1     {  \left\{ \begin{array}{#1}   }
\def\braceend         {  \end{array}\right.   }
\def\parn                 { \par\noindent  }
\def\parbigskip        {  \par\bigskip  }
\def\parmedskip        {  \par\medskip  }
\def\parsmallskip      {  \par\smallskip  }
\def\parbigskipn      { \parbigskip\noindent }
\def\parmedskipn     { \parmedskip\noindent }

\def\parag#1           {\paragraph{#1} \mbox{ }\parmedskip\noindent}
\def\rightfigspacebegin  {  \par\noindent\begin{minipage}[t]{10cm}  }
\def\rightfigspaceend    {  \end{minipage}\par\noindent  }
\def\leftfigspacebegin   {  \par\noindent
                             \hspace*{10cm}\begin{minipage}[t]{6cm} }
\def\leftfigspaceend     {  \end{minipage}\par\noindent  }
\def\titleandfile#1#2   {  \begin{center}{\Large\bf #1}\end{center}
                            \par\begin{flushright} #2 \end{flushright}  }
\def\msection#1      {  \begin{center} \section{#1} \end{center}   }
\def\nsection#1      {  \let\boldface\bf \def\bf{} \section{#1}
                           \let\bf\boldface   }
\def\mnsection#1     {  \begin{center} \nsection{#1} \end{center}  }
\def\capsection#1    {  \let\boldface\bf \def\bf{\sc} \section{#1}
                           \let\bf\boldface   }
\def\mcapsection#1   {  \begin{center} \capsection{#1} \end{center} }
\newcommand{\nullify}[1]{}
\begin{document}
%%%%%%%%%%%%%%%%%%%%%%
%%%%%%%%%%%%%%%%%%%%%%
% lv3-0.tex
%%%%%%%%%%%%%%%%%%%%%%%%
\papertitlepage
%\preprintnumber{DESY 93-043 \\ UT Komaba 93-6  \\ March 1993}
%
\vspace*{0.6cm}
\Title{ On the Exact Operator Formalism of \\ Two-Dimensional
 Liouville Quantum Gravity \\ in Minkowski Spacetime \\}
\Authors{{\sc Yoichi Kazama}\footnote[2]{e-mail address:\quad
kazama@tkyvax.phys.s.u-tokyo.ac.jp} \\
 Institute of Physics, University of Tokyo, \\
 Komaba, Tokyo 153 Japan \\
 \vspace{0.4cm}
 and \\
 \vspace{0.4cm}
{\sc Hermann Nicolai}\footnote[3]{e-mail address:\quad
 i02nic@dhhdesy3.bitnet} \\
 II. Institut f\"ur Theoretische Physik,
  Universit\"at Hamburg \\
 Luruper Chaussee 149, 2000 Hamburg 50, F.R.G.}
\Abstract
\baselineskip 3.5ex
 A detailed reexamination is made of the exact operator formalism
 of two-dimensional Liouville quantum gravity in Minkowski spacetime
 with the cosmological term fully taken into account.
 Making use of the canonical mapping from the interacting Liouville
field into a free field, we focus on the problem of how the
Liouville exponential operator should be properly defined.
In particular, the condition of mutual locality among the  exponential
operators is carefully analyzed, and a new solution, which is
neither smoothly connected nor relatively local to the
existing solution, is found. Our analysis indicates that, in
Minkowski spacetime, coupling gravity to matter with central charge
$d<1$ is problematical. For $d=1$, our new solution appears to be the
appropriate one; for this value of $d$, we demonstrate that the
operator equation of motion is satisfied to all
orders in the cosmological constant with a certain regularization.
As an application of the formalism, an attempt is made to study how
the basic generators of the ground ring get modified due to the
inclusion of the cosmological term.  Our investigation, although
 incomplete, suggests that in terms of the canonically
mapped free field the ground ring is not modified.
\newpage
%%%%%%%%%%%%%%%%%%%%%%%%%%%%%%
\baselineskip 3.5ex
%%%%%%%%%%%%%%%%%%%%%%%%%%%%%%
%%%%%%%%%%%%%%%%%%%%%%
% lv3-1.tex
%%%%%%%%%%%%%%%%%%%%%
\renewcommand{\thefootnote}{\arabic{footnote}}
\setcounter{footnote}{0}
%%%%%%%%%%%%%%%%%%%%%%%
\section{Introduction}
%%%%%%%%%%%%%%%%%%%%%%%
Beginning with  the discovery \cite{MM} that the matrix model
techniques together  with the idea of the double scaling limit
can be used to define and solve  models of two-dimensional Euclidean
gravity (or random surfaces), our understanding of the subject has
been advanced considerably.
 (For review see for example \cite{MMrev} and  references cited
therein.) Despite these successes, however, there still remain a number
 of important issues yet to be clarified.  Notable among them is the
problem of space-time interpretation.  Due to its formulation without
the explicit appearance of the metric, the matrix model offers only a
modest insight into how the gravitational and the matter degrees of
freedom interact.   \par
One possible way to clarify this problem is to study how one can make
contact with the continuum formulation, for example in the conformal
gauge, where the metric degree of freedom appears explicitly as the
Liouville field.  Here the problem is how to treat the notoriously
tricky dynamics generated by the cosmological term $\mu^2\re^\varphi$.
\par
Recently there have been attempts to understand Liouville dynamics
via path-integral formalism with certain amount of success
\cite{GTW}\nocite{Sg}\nocite{BK}\nocite{GL}\nocite{Do}\nocite{Ka}--\cite{AD}.
  In this method one first performs the integration  over
 the Liouville zero mode and observes that if the total
\lq\lq momentum", which is in general fractional, were a positive integer the
 remaining integral becomes
effectively that for a free field with vertex operator insertions
 (although with a formally divergent factor).  Upon performing this
free-field integral one  \lq\lq analytically continues" back to
the original fractional momentum.  The method appears to be non-rigorous
(see however \cite{AD} for justification)
 but it reproduces the matrix model results \cite{FK}
 for up to 3 point functions on a sphere and a torus. \par
Although it has captured an important aspect of the Liouville theory,
the formalism above which integrates out the Liouville zero mode in the
beginning appears to be awkward for unravelling the deep symmetry
structure of the theory. For this purpose, the operator
formalism would appear to be more suitable.
Indeed, the existence of discrete physical states
 \cite{GKN}\nocite{GK}\nocite{DJR}\nocite{DG}\nocite{Pv1}\nocite{LZ}
--\cite{BMP} and the
associated $W_\infty$ algebra \cite{AJ}\nocite{AS}\nocite{GMMMO}
\nocite{MPY}\nocite{MS}--\cite{DDMW} first found in the
matrix model approach have been analyzed in the operator formalism and
such intriguing structures as the \lq\lq ground ring"
 \cite{Wn:GR}\nocite{KP}--\cite{WZ} have been discovered.
These developments have clearly demonstrated the utility
of the operator formalism but there is a caveat: in these
works the Liouville field is treated as a free
field and hence the role of the Liouville dynamics remains unclear. Also
it can be suspected that the naive perturbative treatment of the
exponential interaction may not be justified.
In view of this situation, it is an appropriate time to focus on
the exact operator formalism for the Liouville theory and examine how
it can be applied to the clarification of the aforementioned problems.
\par
Exact operator quantization for the Liouville theory has
been developed over the past ten years
\cite{GN}\nocite{BCT}\nocite{BCGT}\nocite{JM}
\nocite{DJ}\nocite{OW}\nocite{Wt}\nocite{Bn}--\cite{Gs}
(for a review see \cite{Dr}.) In \cite{Gs}, applications of the
operator formalism to two-dimensional quantum gravity coupled to
matter are discussed from a modern point of view, where special emphasis
is placed on the quantum group structure of Liouville theory, and
various results previously obtained by matrix model techniques are
rederived (see also \cite{GsRev} for a recent review).
Despite the fact that many intriguing results have been
obtained in these works, we feel that some subtle issues have been
left unsettled (or overlooked). A major problem, and one that has still
not been satisfactorily resolved in our opinion, concerns the rigorous
construction of exponential Liouville operators of arbitrary weight
having the right conformal properties and satisfying locality.
The purpose of this article is to try to point out these and other
subtleties as clearly as possible and propose solutions to some of these
problems. Because of the technicalities involved we shall not explain
the nature of the subtleties in detail here, but nevertheless let us
mention one novel result of our investigation which may have
serious consequences.  As in the traditional treatment, we shall
perform a canonical operator quantization in Minkowski space by a
non-linear and non-local canonical transformation which maps the
interacting Liouville field into a free field.  In this approach, the
requirement of mutual locality among exponential operators plays a
crucial role. The locality condition was recently reanalyzed by Otto and
Weigt in \cite{OW}. They proposed a new definition of the exponential
operator that is obtained from the usual exponential series by a kind of
quantum deformation, and checked the locality of these operators
in a perturbation expansion in terms of the cosmological
constant $\mu^2$ up to third order\footnote{J. Schnittger has
informed us that he has now verified the locality
condition for the deformed quantum Liouville operator
proposed in \cite{OW} to much higher orders.}. We have carefully
reexamined this locality equation and found an additional solution
(valid to all orders in $\mu^2$) which is not smoothly connected to
the solution of \cite{OW}.  Moreover, our analysis
casts  some doubt (at least in Minkowski space) on the validity of the
conventional treatment of the theory coupled with $d\le 1$ matter.
If we postulate that  the operator $\re^\ve$, which appears in the
Liouville equation of motion,
should belong to the set of admissible (i.e. mutually local)
operators, we find that (i) $d<1$ is not allowed and (ii) for
$d=1$ our solution, not the one constructed in \cite{OW}, is relevant.
Since the notion of locality (or micro-causality) does not enter
in the Euclidean treatment,  our result is not necessarily
in direct conflict with the matrix model, but it certainly calls for
further investigation. \par
With the formalism developed for the $d=1$ case, we then made an attempt
to study how the structure of the ground ring gets modified when
the exponential interaction is turned on. The results we have
obtained so far are unfortunately difficult to interpret. To be sure,
it is straightforward to write down candidates for the basic
generators of the ground ring in terms of the interacting fields with
the expected conformal tranformation properties. However, the
BRST transformation properties are not completely dictated by
the conformal properties alone, and these candidates, unlike their
free field counterparts, turn out not to be BRST invariants. This
seems to indicate that we have exactly the same ground ring generators
as for vanishing cosmological constant. Nonetheless, there is a
large difference with the the $\mu^2 =0$ case. The free field in
our case is {\it not} the Liouville field itself, but the canonically
mapped field which is difficult to interpret physically.  To put it
differently, a simple generator in terms of the  free field is
actually a very complicated one in terms of the Liouville field.
It is not entirely clear to us how our results are related to those of
\cite{Li},\cite{Dots}, where the ground ring with non-vanishing
cosmological constant is also discussed. In \cite{Dots}, the basic
idea is to determine the deformations of the ground ring by
``fusion", i.e. by evaluating operator products of the free field
ground ring operators in a perturbation expansion in $\mu^2$; in this
way one arrives at the conclusion that there appear {\it extra}
operators besides the ones already present in the free field case.
Since we have not computed products of (interacting) ground ring
operators, we cannot ascertain the compatibility of the two approaches
at this point. We note the fact that the expansion in $\mu^2$
of the exact ground ring operators, on which our analysis is based,
breaks off after finitely many terms precisely for the special
weights which appear in the associated Liouville exponentials.
It would be interesting to see whether this observation is related
to the fact that only finitely many terms remain in the perturbative
expansion of \cite{Dots} due to charge screening.
Obviously, more work is needed to clarify these issues.
\par
Despite a long history, pedagogical expositions of the operator
approach to Liouville theory seem to be scarce. For this reason,
we have decided to organize  our paper in such a way that it
can (hopefully) also serve as a relatively self-contained review.
We start in section 2 with a review of the classical properties
of the Liouville theory.  In particular, we provide a rather detailed
exposition of the non-linear and non-local canonical transformation
which maps the interacting Liouville field into a free field.
With the use of this canonical tranformation, we proceed to the
operator quantization of the theory in section 3.  The central issue
will be the question of the proper definition of exponential operators
and their locality properties.  We will also demonstrate that
the operator Liouville equation of motion is satisfied to all orders
 in $\mu^2$ for certain set of values for the matter central charge
 including $d=1$.  In section 4 we shall describe an attempt to construct
 the basic generators of the ground ring including the cosmological
term.  Finally in section 5 we will give discussions on the issues
which are left unsolved in this work. An appendix is provided to discuss
the conversion between Minkowski and Euclidean formulations.
%%%%%%%%%%%%%%%%%%%%%%%%%%%%%%%%%%%%%%%%%%%%%%%%%%%%%
%%%%%%%%%%%%%%%%%%%%%%%%
%  lv3-2.tex
%%%%%%%%%%%%%%%%%%%%%%%%%%%%%%%%%%%%%%%%%%%%%%%%%%%%%%%%
\section{Classical Properties of Liouville Theory}
%%%%%%%%%%%%%%%%%%%%%%%%%%%%%%%%%%%%%%%%%%%%%%%%%%%%%%%%%
\Subsection{General Remarks}
%%%%%%%%%%%%%%%%%%%%%%%%%%%%%%%%%%
In this section we describe some classical properties of Liouville
theory \cite{GN}--\cite{Bn}; as for our conventions and notation
we will mostly follow \cite{OW}.
Let us briefly recall how the Liouville action arises
in string theory, and why it plays an important role there.
Any conformal field theory can be formulated as a two-dimensional
field theory of matter fields coupled to a gravitational background,
which is described by a metric tensor $g_{\m \n}$. In two dimensions,
$g_{\m \n}$ carries no propagating but only topological degrees
of freedom. More specifically, it can always be brought to the form
\eqabegin
g_{\m \n} =  \re^\ve \hat g_{\m \n}
\eqaend
by means of two-dimensional diffeomorphisms. Here, the background
metric  $\hat g_{\m \n}$ locally can be chosen equal to the
flat Minkowski metric (i.e. in each coordinate patch, we can set
$\hat g_{\m \n} = \eta_{\m \n}$). Globally, it will depend
on the moduli characterizing the conformal equivalence class
of the world sheet; we will, however, ignore these topological degrees
of freedom in most of the remainder. The factor multiplying
the background metric $\hat g_{\m \n}$ will be referred
to as the ``conformal factor" in the sequel, and $\ve$ is the
Liouville field. In a conformally invariant theory, the classical
Lagrangian is not only invariant under two-dimensional diffeomorphisms,
but in addition invariant under Weyl rescalings of the metric. This
invariance allows us to gauge the conformal factor to unity, and
the Liouville field therefore decouples from the matter sector.
However, as is well known \cite{Pv2}, this decoupling in general
cannot be maintained at the quantum level, because integration over
the matter fields induces the quantum action
\eqabegin
S = {1\over {\g^2}}\int d\ta d\s \sqrt{\hat g} \left(
     \ft12 \hat g^{\m \n} \del_\mu \ve \del_\n \ve  -
  \m^2 \re^\ve  - Q \ve \hat R \right)  \label{action1}
\eqaend
or, with $\hat g_{\m \n} = \eta_{\m \n}$,
\eqabegin
S = {1\over {\g^2}}\int d\ta d\s \left( \ft12 \del_\mu \ve \del^\mu \ve
     - \m^2 \re^\ve \right)\comma \label{action2}
\eqaend
where $\m^2$ is the cosmological constant, $Q$ the background charge.
Furthermore, we have
\eqabegin
{1\over {\g^2}} \equiv {{26-d}\over {48\pi}}\period
\eqaend
$d$ is the number of scalar matter fields \footnote{For definiteness,
we consider only bosonic theories.} or, more generally, the total
central charge of the matter sector (which need not even be
described by a Lagrangian). Consequently, at the quantum level,
the Liouville field decouples only for $d=26$, which is the critical
dimension for a bosonic target space. For $d< 26$, on the other
hand, the Liouville field $\ve$ must be taken into account as an
extra degree of freedom, whose dynamics is governed by the
non-polynomial action (\ref{action1}) \footnote{The case $d>26$ is not of
interest as there are always negative norm states in the physical
Hilbert space.}. In order to make sense out of such
a ``subcritical" theory, one is thus forced to study the Liouville
action in the context of two-dimensional quantum field theory.
The hope, then, is that through its non-linear dynamics, the
Liouville field adjusts its contribution to the conformal anomaly
in such a way that conformal invariance is restored in the full theory,
and thereby a consistent theory can be formulated also outside the
critical dimension. We remark that, in many approaches, the
(renormalized) cosmological constant is set to zero, so the
complicated dynamics of the Liouville field is masked by an
effective free theory. However, we will not adopt this point of view
and assume $\m^2 \neq 0$ in this paper.

One of the central problems of Liouville theory is the construction
of quantum operators corresponding to exponentials of the Liouville
field, i.e. $\re^{\l \ve}$ for arbitrary weights $\l$. We will discuss
this problem in much detail in section 3.2, but let us emphasize
already at this point that these operators are indispensable for
the construction of physical vertex operators. The main reason is
that ordinary string vertex operators built out of matter fields
only will usually fail to have the correct conformal dimension, and
must be ``dressed up" by suitable powers of the worldsheet
metric and its determinant before they can be consistently
integrated over the world sheet. Since the conformal dimension of
$\re^{\l \ve}$ depends upon $d$, the conformal dimensions of
the matter field vertex operators will also have to be modified
away from the critical dimension.

Before moving on to describe the classical aspects of Liouville
theory, we note another important point. Due to quantum effects,
we will have to replace the coupling constant multiplying the
action (\ref{action1}) by the renormalized value \cite{GN}
\eqabegin
\left( {1\over {\g^2}} \right)_{ren}
      \equiv {{25-d}\over {48\pi}} = {1\over {16\pi \hbar}}\period
  \label{ffactor}
\eqaend
The shift from 26 to 25 is a quantum effect and will be
explained below. In the above equation, we have also
introduced  ``Planck's constant" $\hbar \equiv {3\over{25-d}}$.
This terminology has the advantage that we can think of the
limit $d \rightarrow - \infty$ as some kind of semi-classical limit
in which the theory simplifies considerably.
%%%%%%%%%%%%%%%%%%%%%%%%%%%%%%%%%%%%%%%%%%
\Subsection{General Solution}
%%%%%%%%%%%%%%%%%%%%%%%%%%%%%%%%%%%%%%%%%%%%%%
The classical equation of motion following from the action
(\ref{action2}) reads
\eqabegin
4 \del_+ \del_- \ve + \m^2 \re^\ve = 0 \comma \label{eqmotion}
\eqaend
where, for later convenience, we have introduced the light-cone
coordinates
\eqabegin
\xi^\pm \equiv \ta \pm \s \qquad
\del_\pm \equiv \ft12 (\del_\ta \pm \del_\s ) \;\;\; \Rightarrow \;\;\;
    \del_\ta^2 - \del_\s^2 = 4 \del_+ \del_- \period
\eqaend
(\ref{eqmotion}) is a non-linear partial differential equation, whose
general
solution has been known for a long time (and was, in fact, known
to Liouville \cite{Le}). It is most conveniently expressed in
terms of left and right moving waves, i.e. two arbitrary
functions $A = A (\xip )$ and $B = B (\xim )$:
\eqabegin
\ve (\ta , \s ) = \log \left( {8\over {\m^2}}
{{\del_+ A (\xip ) \del_- B (\xim )} \over
 {\left(1 + A(\xip ) B(\xim )\right)^2}} \right) \period \label{solution}
\eqaend
Observe that this representation is not unique and differs from the
one used in earlier papers \cite{GN}. The above parametrization was
chosen by \cite{DJ}\cite{OW}, and will turn out to be the most convenient
for our subsequent considerations. The fact that the above
equation can be solved exactly may be regarded as a consequence
of the integrabiliy of Liouville theory, which follows from the
existence of a Lax pair \cite{Lpair}.
Periodicity in the space coordinate requires the
boundary condition $\ve (\ta , \s + 2\pi ) = \ve (\ta , \s )$.
This does not imply that the ``constituent functions" $A,B$ are
themselves periodic; rather, the periodicity constraint on the
Liouville field is compatible with
$SL(2,{\bf R})$ (i.e. M\"obius) transformations on $A$ and $B$, viz.
\eqabegin
A(\xip + 2\pi ) ={{\a A (\xip ) + \b}\over {\g A(\xip ) + \d}}
\comma \qquad
B(\xim - 2\pi ) ={{\d B (\xim )-\g}\over {-\b B(\xim ) + \a}} \comma
\label{bc}
\eqaend
where $\a \d - \b \g = 1$. Note that $B$ transforms ``contragrediently"
to $A$; this is the reason why some authors prefer to work with the
field $-B(\xim )^{-1}$ instead, which transforms like $A$.
In choosing boundary conditions below, we will not make use
of the full M\"obius group, but rather put $\b = \g =0$.
The solution (\ref{solution}) can be equivalently expressed as
\eqabegin
\ve (\ta , \s ) = \log \del_+ A (\xip ) + \log \left({8\over {\m^2}}
  \del_- B(\xim )\right) - 2 \log \left( 1 + Y(\ta , \s )\right)
  \comma \label{phi}
\eqaend
where
\eqabegin
Y(\ta , \s ) \equiv A(\xip ) B(\xim )
\eqaend
is the only part of $\ve$ mixing left and right movers.

Having exhibited the explicit solution in terms of left and right
movers, the next task is to construct a free field $\p$ out of them.
In view of the different parametrizations alluded to before,
the identification of the free field is not unambiguous. We will
here follow \cite{DJ} and \cite{OW} and define $\p$ by
\eqabegin
\p (\ta , \s) = \p^+ (\xip ) + \p^- (\xim ) \equiv \log \del_+ A(\xip )
  + \log \left({8\over {\m^2}} \del_- B(\xim ) \right)
\eqaend
so that
\eqabegin
\ve = \p - 2 \log (1+Y) \period \label{phi2}
\eqaend
These equations will serve as our basis for all subsequent
considerations, in particular those concerning the quantum theory.

Integration of the differential equations
$\del_+ A = \re^{\p^+}$ and $\del_- B =(\mu^2 /8) \re^{\p^-}$ is
straightforward.
Assuming the simplified boundary conditions
$A(\xip + 2\pi )= \a A(\xip )$ and $B(\xim -2\pi )= \a^{-1} B(\xim )$
we obtain \cite{OW}
\eqabegin
A(\xip ) &=& c(\a )\int_0^{2\pi} d\xi' \, E_\a (\xip - \xi' )
           \re^{\p^+ (\xi')} \comma \label{A}  \\
B(\xim ) &=&  {{\m^2}\over 8} c(\a )
       \int_0^{2\pi}d\xi'' \, E_\a (\xim -\xi'')
           \re^{\p^- (\xi'')} \comma
\eqaend
where
\eqabegin
c(\a )\equiv
   \left( \sqrt{\a} -{1\over {\sqrt{\a}}}\right)^{-1} \comma
\eqaend
and
\eqabegin
E_\a (\xi ) \equiv \exp \left( \ft12 \log \a \e (\xi ) \right)
= E_{1\over {\a}} (-\xi )\comma
\eqaend
with $\e (\xi )$ the usual staircase function, i.e. $\e (\xi ) =
2n+1$ for $2n\pi < \xi < 2(n+1)\pi$. When written out, (\ref{A}) reads
\eqabegin
A(\xi^+ ) = c(\a ) \left( \int_0^{\xi^+} d\xi'
                \re^{{1\over 2} \log \a}
\re^{\p^+ (\xi'  )}+ \int_{\xi^+}^{2\pi} d\xi' \re^{-{1\over 2} \log \a}
   \re^{\p^+ (\xi' )}  \right) \period
\eqaend
The function $E_\a$ (which is actually a distribution) obeys
\eqabegin
{\del \over {\del \xi }}E_\a (\xi )= {1 \over {c(\a )}}\, \d (\xi )
\comma \label{delE}
\eqaend
where $\d (\xi )$ is the $2\pi$-periodic delta-function.
Note that we cannot choose the functions $A,B$ to be simply
periodic because $c(\a)$ diverges at $\a =1$. Solving for
$Y(\ta , \s )$, we obtain
\eqabegin
Y(\ta , \s ) = {{\m^2}\over 8}
   c(\a )^2 \int_0^{2\pi} d\s' \int_0^{2\pi} d\s'' \,
E_\a (\s - \s') E_{1\over \a} (\s- \s'' ) \re^{\p^+ (\ta + \s' )}
\re^{\p^- (\ta - \s'')} \period \label{Y}
\eqaend
It is now easy to check that $Y$ is a periodic function of $\s$,
although $A$ and $B$ are not periodic separately (note that with
the more general boundary condition (\ref{bc}) $Y$ would not
have been periodic). Since, for the
canonical formalism we will be mostly concerned with equal time
expressions, we have explicitly displayed the dependence on the
time and space variables in (\ref{Y}); this will permit us to put
$\ta =0$ wherever appropriate.
Although we will not make use of this fact, we note that $Y(\ta , \s)$
can also be represented directly in terms of the free field $\p$ and
its time derivative $\dot \p$ at time $\ta$. The resulting expression
is local in $\tau$, but non-local in the space variable $\s$.

Given the result for the Liouville field as a non-local function of
the free field $\p$, we can now define exponentials with arbitrary
weights $\l$ through the expansion
\eqabegin
\re^{\l \ve} = e^{\l \p} \left( 1 + Y \right)^{-2\l} =
\re^{\l \p } \sum_{n=0}^{\infty} {{(-1)^n}\over {n!}}
{{\G (2\l +n)}\over {\G (2\l )}} Y^n \period \label{explphi1}
\eqaend
(We have displayed the expression appropriate for positive $\lambda$ )
Observe that this is actually an expansion in powers of the
cosmological constant, as $Y$ is proportional to $\m^2$ by (\ref{Y}).
Note also that when $-2\lambda$ is a non-negative integer the
expansion contains only a finite number of terms.

In the remainder, we will also make use of the Fourier expansion of
the free field $\p$ in terms of oscillators. For the left and
right moving fields $\p^\pm$ this expansion takes the form
\eqabegin
\p^\pm (\xi^\pm ) = \ft12 \g Q + {\g \over {4\pi}} P \xi^\pm
+ {{i\g}\over{\sqrt{4\pi}}} \sum_{n\ne 0} {1\over n} a_n^{(\pm)}
  \re^{-in\xi^\pm} \comma \label{psiexp}
\eqaend
where the extra factor of $\g$ has been inserted so as to obtain
the canonically normalized Poisson bracket relations for $Q,P$ and the
oscillators $a_n^{(\pm)}$.
Comparison with (\ref{phi}) leads to the identification
\eqabegin
\log \a = \ft12 \g P \period \label{alpha}
\eqaend
We have already remarked above that we cannot put $\a =1$ in the
expressions for the interacting Liouville field. Now we can
understand this result in a more physical way: $\a \neq 1$ is
equivalent to $P\neq 0$, i.e. Liouville theory
does not possess a translationally invariant groundstate; of course,
this problem becomes apparent only for $\m^2 \neq 0$. There is
another, and related, difficulty here. Any shift of the Liouville
field by a constant can be absorbed into the cosmological constant,
which is therefore an essentially undetermined parameter. This, in
turn, means that there is no way in which (\ref{explphi1}) can be
thought of as
an expansion in a small parameter. In the absence of a better method,
we will proceed in spite of this difficulty.
%
%%%%%%%%%%%%%%%%%%%%%%%%%%%%%%%%%%%%%%%%
\Subsection{Canonical Transformation}
%%%%%%%%%%%%%%%%%%%%%%%%%%%%%%%%%%%%%%%%
The Poisson brackets are determined from the action (2.1.3).
The equal time brackets are given by
\eqabegin
\{ \ve (\ta , \s )\, , \,  \ve (\ta , \s' )  \} =
\{ \dot \ve (\ta , \s )\,  , \, \dot \ve (\ta , \s' )  \} = 0
\qquad {\rm for} \;\;\; \s \neq \s' \comma\label{PB1}
\eqaend
\eqabegin
\{ \ve (\ta , \s )\, ,\,  {1\over {\g^2}} \dot \ve (\ta , \s' )  \} =
    \d (\s - \s' )\comma\label{PB2}
\eqaend
where the dot denotes the derivative with respect to $\ta$.
The factor in front of $\dot \ve$ is just the coupling constant
multiplying the action (\ref{action2}).

It is quite remarkable that the non-local transformation connecting
the Liouville field $\ve$ and the free field $\p$ is canonical
\cite{GN},\cite{BCT}. In the present formulation, this means that
the above brackets are equivalent to
\eqabegin
\{ \p (\ta , \s )\, ,\, \p (\ta , \s' )  \} =
\{ \dot \p (\ta , \s )\, ,\, \dot \p (\ta , \s' )  \} = 0
\qquad {\rm for} \;\;\; \s \neq \s' \comma
\eqaend
\eqabegin
\{ \p (\ta , \s )\, ,\, {1\over {\g^2}} \dot \p (\ta , \s' )  \} =
    \d (\s - \s' ) \period
\eqaend
The proof of this assertion requires a rather tedious computation.
Since our setup, and in particular, our definition of the free field
in (2.2.7), is different from the one in \cite{GN}, \cite{BCT},
we present the details of this proof here.
We have found that the simplest method is to proceed ``backwards"
by showing that the free field Poisson brackets imply the ones
in terms of $\ve$ above. We now give some useful intermediate
relations. Since all of these are equal time commutators, we will omit
the time coordinate in the formulas listed below; however, we alert
the reader that in all relations involving time derivates, we must keep
$\ta$ as a variable at the intermediate stages of the calculation
and can put $\ta =0$ only at the very end.
Also one must not forget that $\alpha$ contains the zero mode $P$ and
hence has non-trivial bracket with $\psi$, which contains $Q$.
A prime $'$ denotes
the derivative with respect to the spatial coordinate $\s$. \par
Let us first list the brackets between the free field $\p$ and
the quantities $A,B$; to make the formulas less cumbersome, we put
$\g^2 =0$ in eqs. (2.3.5--21). So we obtain
\eqabegin
\{ \p(\s_1 ) \, ,\, A(\s_2) \} &=&
{{\s_1}\over {4\pi}} A(\s_2 ) - {{c(\a )}\over 2} A(\s_1 )
      E_{{1\over \a}} (\s_1 - \s_2 )   \nn \\
  && \quad - {1\over 4} A(\s_2 ) \e (\s_1 - \s_2 )\comma
\eqaend
\eqabegin
\{ \p(\s_1 ) \, ,\, A'(\s_2) \} =
{{\s_1}\over {4\pi}} A'(\s_2 ) - {1\over 4} A'(\s_2 ) \e (\s_1 - \s_2 )
\comma
\eqaend
\eqabegin
\{ \dot \p(\s_1 ) \, ,\, A(\s_2) \} =
 - {{c(\a )}\over 2}  A'(\s_1 ) E_{{1\over \a}} (\s_1 - \s_2 ) \comma
\eqaend
\eqabegin
\{ \dot \p(\s_1 ) \, ,\, A'(\s_2) \} =
- \half  A'(\s_2 ) \d (\s_1 - \s_2 )\comma
\eqaend
\eqabegin
\{ \p(\s_1 ) \, ,\, B(\s_2) \} &=& -
{{\s_1}\over {4\pi}} B(\s_2 ) - {{c(\a )}\over 2}
      E_{{1\over \a}} (\s_2 - \s_1 ) B(\s_1 )  \nn \\
  &&\quad  + {1\over 4} \e (\s_1 - \s_2 ) B(\s_2 )\comma
\eqaend
\eqabegin
\{ \p(\s_1 ) \, ,\, B'(\s_2) \} =  -
{{\s_1}\over {4\pi}} B'(\s_2 )-{1\over 4} \e (\s_1 -\s_2 ) B'(\s_2 )
\comma
\eqaend
\eqabegin
\{ \dot \p(\s_1 ) \, ,\, B(\s_2) \} =
 + {{c(\a )}\over 2} E_{{1\over \a}} (\s_2 - \s_1 ) B'(\s_1 )\comma
\eqaend
\eqabegin
\{ \dot \p(\s_1 ) \, ,\, B'(\s_2) \} =
 -  \half B'(\s_1 ) \d (\s_1 - \s_2 )\period
\eqaend
{}From these brackets, we derive
\eqabegin
\{ \p (\s_1 )\,, \, Y(\s_2) \} &=& - \half c(\a ) \left( A(\s_1 )
  E_{{1\over \a}} (\s_1 - \s_2 ) B (\s_2 ) +
  A(\s_2 ) E_{{1\over \a}} (\s_2 - \s_1 ) B (\s_1 ) \right)  \nn  \\
  &=& \{ \p (\s_2 )\, ,\, Y (\s_1 ) \} \comma
\eqaend
\eqabegin
\{ \dot \p (\s_1 )\, ,\,  Y(\s_2) \} &=& - \half c(\a ) \left( A'(\s_1 )
  E_{{1\over \a}} (\s_1 - \s_2 ) B (\s_2 ) -
  A(\s_2 ) E_{{1\over \a}} (\s_2 - \s_1 ) B' (\s_1 ) \right)  \nn  \\
  &=& \{ \p (\s_2 )\, ,\, \dot Y (\s_1 ) \} \comma
\eqaend
\eqabegin
\{ \p (\s_1 )\, ,\, \dot Y(\s_2) \} &=&  \half c(\a ) \left( A(\s_1 )
  E_{{1\over \a}} (\s_1 - \s_2 ) B' (\s_2 ) -
  A'(\s_2 ) E_{{1\over \a}} (\s_2 - \s_1 ) B (\s_1 ) \right)  \nn  \\
  &=& \{ \dot \p (\s_2 )\, ,\, Y (\s_1 ) \}\comma
\eqaend
\eqabegin
\{ \dot \p (\s_1 ) \,, \,  \dot Y (\s_2) \} &=&
     -\half \d (\s_1 - \s_2 ) \dot Y (\s_2 )    \\
          &+&  \half c(\a ) \left( A'(\s_1 )
  E_{{1\over \a}} (\s_1 - \s_2 ) B' (\s_2 )  +
  A'(\s_2 ) E_{{1\over \a}} (\s_2 - \s_1 ) B' (\s_1 ) \right)
\period \nn
\eqaend
The equal time relations involving commutators of $Y$ and $\dot Y$ read
\eqabegin
\{ Y(\s_1 )\, ,\,  Y(\s_2 )\} &=& \half \left( Y(\s_1 )- Y(\s_2 )\right)
    \{ \p (\s_1 )\, ,\,  Y(\s_2 ) \} \nn  \\
     &=& \half \left( Y(\s_1 )- Y(\s_2 ) \right)
    \{ \p (\s_2 )\, ,\,  Y(\s_1 ) \} \comma
\eqaend
\eqabegin
\{ Y(\s_1 )\, ,\,  \dot Y (\s_2 ) \} =
 &-& \half \{ \p (\s_1 )\, ,\, Y(\s_2 ) \} \dot Y (\s_2 ) \nn  \\
 &+& \half \left( Y(\s_1 ) - Y(\s_2 ) \right)
    \{ \p (\s_1 )\, ,\,  \dot Y(\s_2 ) \} \comma
\eqaend
\eqabegin
\{ \dot Y(\s_1 )\, ,\, \dot Y (\s_2 ) \} =
 &-& \half \{ \p (\s_2 )\, ,\, \dot Y(\s_1 ) \} \dot Y (\s_2 ) +
 \half \{ \p (\s_1 )\, ,\,  \dot Y(\s_2 ) \} \dot Y (\s_1 ) \nn  \\
 &+& \half \left( Y(\s_1 ) - Y(\s_2 ) \right)
    \{ \dot \p (\s_1 )\, ,\,  \dot Y(\s_2 ) \} \period
\eqaend
These results can now be inserted to rewrite (\ref{PB1}) and
 (\ref{PB2}) in terms of free field commutators. For instance, we get
\eqabegin
\{ \ve (\s_1 ) \, , \, \ve (\s_2 ) \} &=&
 - {2 \over {1+Y(\s_2 )}} \{ \p (\s_1 )\, ,\, Y(\s_2 ) \}
 + {2 \over {1+Y(\s_1 )}} \{ \p (\s_2 )\, ,\, Y(\s_1 ) \} \nn \\
&&  + {4\over{(1+Y(\s_1))(1+Y(\s_2))}} \{ Y(\s_1 ) \, , \, Y(\s_2 ) \}
\period
\eqaend
The last term can be rewritten by use of (2.3.17)
\eqabegin
{{Y(\s_1 ) - Y(\s_2 )}\over {(1+Y(\s_1))(1+Y(\s_2 ))}}
\Big( \{ \p (\s_1 ) \, ,\, Y(\s_2 )\}
 +    \{ \p (\s_2 ) \, ,\, Y(\s_1 )\} \Big)    \nn
\eqaend
\eqabegin
 = {2 \over {1+Y(\s_2 )}} \{ \p (\s_1 )\, ,\, Y(\s_2 ) \}
 - {2 \over {1+Y(\s_1 )}} \{ \p (\s_2 )\, ,\, Y(\s_1 ) \} \period
\eqaend
and thus precisely cancels the first two terms. The proof
of the remaining Poisson brackets along these lines is now
completely analogous.

By standard arguments it can be shown that the free field
Poisson brackets are equivalent to the following brackets for the
oscillators
\eqabegin
\{ Q,P\}=1 \;\;\;, \,\,\, i\{ a_m^{(+)} , a_n^{(+)} \} =
 i\{ a_m^{(-)} , a_n^{(-)} \} = m \d_{m+n,0} \period
\eqaend
(The brackets that have not been listed vanish). \par
Although the proof of the canonical nature of the transformation
is already complete, it is instructive to display explicitly
how the Hamiltonian gets transformed. The Hamiltonian following from
the action (\ref{action2}) is given by
\eqabegin
 H &=& \int d\sigma \left( \half \gamma^2 \Pi^2_\varphi
      + \half {1\over \gamma^2} {\varphi'} ^2
     +{\mu^2 \over \gamma^2} \re^\varphi \right) \nn \\
   &=& {1\over \gamma^2} \int d\sigma \left( \half \dot{\varphi} ^2
      + \half {\varphi'} ^2
     +\mu^2  \re^\varphi \right)  \comma
\eqaend
where $  \Pi_\varphi = {\dot{\varphi} \over \gamma^2} $ is the canonical
 momentum.  We now substitute the expression (\ref{phi2}) and make
use of identities such as
\eqabegin
 \delplus\psi \delplus Y &=& \delplus^2 Y \comma \\
 \delminus \psi \delminus Y &=& \delminus ^2 Y \comma \\
 \delplus\delminus Y &=& {\mu^2 \over 8} \re^\psi \comma \\
 \delplus Y \delminus Y  - Y \delplus\delminus Y &=& 0 \comma
 \eqaend
which easily follow from the definitions of $Y$ and $\psi$.  After
some calculations we obtain
\eqabegin
  H &=& {1\over \gamma^2} \int_0^{2\pi}
 d\sigma \left[ \half\left( \dot{\psi}^2
+ {\psi'}^2 \right) -4 \del_\sigma ^2 \ln (1+Y)\right]  \nn\\
 &=& {1\over \gamma^2} \int_0^{2\pi} d\sigma
 \half\left( \dot{\psi}^2 + {\psi'}^2 \right) \period
\eqaend
This is nothing but a Hamiltonian for a free field $\psi$.  So there
are two ways to look at the Liouville dynamics:  one can say that the
Liouville field evolves in a complicated way because of the exponential
interaction.  Alternatively one may say that the Liouvile dynamics
 is non-trivial because the Liouville field is a complicated
combination of a free field which evolves trivially.
%%%%%%%%%%%%%%%%%%%%%%%%%%%%%%%%%%%%%%%%%%%%%%%%%%%%%%%
\Subsection{Energy-Momentum Tensor and Conformal Transformations}
%%%%%%%%%%%%%%%%%%%%%%%%%%%%%%%%%%%%%%%%%%%%%%%%%%%%%%%%%
The energy momentum tensor is defined by
\eqabegin
T_{\m \n} = 4\pi {{\d S}\over {\d \hat g^{\m \n}}}\period
\eqaend
It is not difficult to see that the trace of $T_{\m \n}$ is
proportional to the Liouville equation of motion and therefore
vanishes on-shell. It is in this sense that we can regard Liouville
theory as a conformal field theory, although the basic Liouville field
does not decompose into a sum of left and right movers.
Switching to light-cone notation, we find for
the remaining traceless components
\eqabegin
T_{\pm \pm} = 4\pi {{\d S}\over {\d \hat g^{\pm \pm}}}=
  {{2\pi}\over {\g^2}}
    \big(  \del_\pm \ve \del_\pm \ve - 2\del_\pm^2 \ve \big)\period
\eqaend
Note that we must keep the fiducial metric $\hat g_{\m \n}$
in order to get the improvement term, and can equate it with the
 flat Minkowski metric only at the very end of the calculation.
{}From conservation of the energy monentum tensor (i.e. $\del_-
T_{++} + \del_+ T_{-+} =0$) and its tracelessness
(i.e. $T_{+-} = 0$), we expect $T_{++}$
($T_{--}$) to depend only on $\xip$ ($\xim$). This is borne out
by an explicit computation, which yields
\eqabegin
 T_{++} ={{2\pi}\over {\g^2}}
    \big(  \del_\pm \p \del_\pm \p -2\del_\pm^2 \p  \big)\period
\eqaend
Therefore, despite the complicated dependence of the Liouville
field on $\ta$ and $\s$, the $++$ and $--$ components of the energy
momentum tensor take a rather simple form in terms of free fields,
being equivalent to those of a free field theory, apart from the
improvement term containing second derivatives. In other words,
as far as its energy momentum tensor is concerned, Liouville theory
behaves very much like a free field theory (with a background charge).

The Virasoro generators $L_m^{(\pm)}$, defined from the expansion
\eqabegin
T_{\pm \pm } (\xi^\pm ) = \sum_m L_m^{(\pm)} \re^{-im\xi^\pm}
\eqaend
are given by (cf. (\ref{psiexp}))
\eqabegin
L_m^{(\pm )} = \ft12 \sum_n a_n^{(\pm )} a_{m-n}^{(\pm )}
         + {{\sqrt{4\pi}}\over{\g}} i m a^{(\pm)}_m \period \label{Lm}
\eqaend
In the sequel, we will omit the superscript $(\pm)$ if there is no
danger of confusion. By use of the canonical brackets given in the
preceding section, one recovers the Virasoro algebra
\eqabegin
i \{ L_m , L_n \} = (m-n) L_{m+n} + {{4\pi}\over {\g^2}}
                 m^3 \d_{m+n,0}\period \label{Viralg}
\eqaend
Consequently, there is a central term already at the classical level.
Note the the anomaly term proportional to $m^3$ can be
converted into the more usual form
$m(m^2-1)$ through the replacement of $m a_m^{(+)}$ by
$(m+1) a_m^{(+)}$ in (\ref{Lm}).

At this point, we can explain the shift from 26 to 25
mentioned in section 2.1. As is well known, in the quantum theory,
the expression for the Virasoro generators must be normal ordered
if it is to be remain well defined. This results in an extra
contribution of
$\ft1{12}$ to the central charge. Since we want the total
central charge to be given by 26, we have to readjust its classical
value in such a way that
\eqabegin
{c\over{12}} = {1\over{12}} + \left( {{4\pi}\over{\g^2}}\right)_{ren}
             = {{26-d}\over{12}}\comma \label{ccharge}
\eqaend
which leads to the renormalized value given in (\ref{ffactor}). We will
henceforth drop the subscript ``$ren$", and always assume the
renormalized value for the coupling constant.
%%%%%%%%%%%%%%%%%%%%%%%%%%%%%%%%%%%%%%%%%%%%%%%%%%%%
%%%%%%%%%%%%%%%%%%%%%%%%%%%%
% lv3-3.tex
%%%%%%%%%%%%%%%%%%%%%%%%%%%%%%%%%%%%%%%%%%
\section{Quantization of Liouville Theory}
%%%%%%%%%%%%%%%%%%%%%%%%%%%%%%%%%%%%%%%%%%%%
\Subsection{Free Field Quantization}
%%%%%%%%%%%%%%%%%%%%%%%%%%%%%%%%%%%%%%%%%%%%
The main result of the foregoing chapter was the demonstration that
the interacting Liouville field can be canonically reexpressed
in terms of the free field $\p$ defined in (2.2.7). In quantizing
Liouville theory, one tries to exploit this equivalence by performing
the quantization in terms of this free field. In this subsection,
we set up our conventions and notation and collect some
basic (and well known) formulas for free fields.

First replace Poisson brackets by commutators in the usual fashion
so that
\eqabegin
[ Q,P ]=i \;\;\;, \,\,\, [ a_m^{(+)} , a_n^{(+)} ] =
 [ a_m^{(-)} , a_n^{(-)} ] = m \d_{m+n,0}
\eqaend
(again, commutators that have not been listed vanish).
This can be inserted into (\ref{psiexp}) for the free field $\p$
to derive the quantum commutators
\eqabegin
[ \p (\ta , \s ) \, ,\, \p (\ta , \s' )  ] =
[  \dot \p (\ta , \s )\, ,\, \dot \p (\ta , \s' ) ] = 0
\qquad {\rm for} \;\;\; \s \neq \s' \comma
\eqaend
\eqabegin
[ \p (\ta , \s )\, ,\, {1\over {\g^2}}\dot \p (\ta ,\s' ) ]
 = i  \d (\s - \s' ) \period
\eqaend
For the derivation of short distance expansions and the
computation of vacuum expectation values it will be convenient to
split the field $\p$ into creation and annihilation parts by defining
\eqabegin
\p^{\pm}_a (\xi^\pm )\equiv {{i\g}\over {\sqrt{4\pi}}} \sum_{n=1}^\infty
  {1\over n} a_n^{(\pm)} \re^{-in\xi^\pm} \comma \label{psia}
\eqaend
and
\eqabegin
\p^{\pm}_c (\xi^\pm )
    \equiv - {{i\g}\over {\sqrt{4\pi}}} \sum_{n=1}^\infty
  {1\over n} a_{-n}^{(\pm)} \re^{+in\xi^\pm}\comma \label{psic}
\eqaend
where zero mode terms involving $Q$ and $P$ have been left out.

All relevant formulas can now be derived from the
reordering relation (for the non-zero modes)
\eqabegin
 \re^{a \psi^+_a(\xip_1)} \re^{b \psi^+_c(\xip_2)} &=&
   \re^{b \psi^+_c(\xip_2)} \re^{a \psi^+_a(\xip_1)} \nn  \\
   && \cdot \mid 1- \re^{i\mid \xip_1 -\xip_2 \mid }
       \mid^{-ab {{\g^2}\over {4\pi}}}
   \cdot  \re^{-i\pi ab \gp \e (\xip_1 -\xip_2) }
  \re^{iab \gp (\xip_1-\xip_2)}\comma \label{exchange}
\eqaend
and the corresponding formula for exponentials of $\p^- (\xim )$,
which reads exactly the same. The relation for the zero modes is
\eqabegin
\re^{aQ} f(P) = f(P+ia) \re^{aQ} \period
\eqaend
To establish these relations, we need the formula
\eqabegin
 \sum_{n\ge 1} {1\over n} \re^{-in(\theta-i\epsilon)}
 &=& -\ln \left( 1-\re^{-i(\theta -i\epsilon)}\right) \nn\\
 &=& -\ln \mid 1-\re^{-i\theta} \mid - i {\pi \over 2}\epsilon(\theta)
   +i{\theta \over 2}\comma
\eqaend
which is obtained by carefully looking at the phase of the logarithm in
the region where $\theta$ is small. In deriving the reordering
relations above, one encounters infinite sums which fail to converge
without regularization. The sums are regularized by adding a small
imaginary part to the argument $\t$;
of course, $\e$ must be taken to zero at the end of the calculation
(this shift is very similar to the well known $i\e$-prescription
needed to make Feynman integrals well defined). The phase factors
will be especially important in our analysis of
the locality condition in section 3.3.

Normal ordering is now defined in the usual fashion by putting
all annihilation operators to the right, except that in the zero mode
sector we use the {\it symmetric} normal ordering given by
\eqabegin
:\re^{aQ} f(P): \equiv  \re^{\half a Q} f(P) \re^{\half a Q} \period
\eqaend
In accordance with these remarks, the fully normal ordered exponential
is now given by
\eqabegin
 :\exp \left( a \p (\ta , \s ) \right): &=&
           \exp \left( \ft12 a \g Q) \right)
     \exp \left( {{a \g}\over{4\pi}} P \ta  \right)
    \exp \left( \ft12 a \g Q  \right)
 \exp \left( a \p^+_c (\ta + \s )\right)    \nn  \\
 && \cdot \exp \left( a \p^+_a (\ta + \s \right)  \exp \left( a \p^-_c
  (\ta - \s )\right) \exp \left( a \p^-_a (\ta - \s ) \right)\period
  \nn\\ &&
\eqaend

By means of the reordering relation (3.1.6), it is straightforward
to prove that
\eqabegin
:\re^{ a\p^+ (\xip_1 )}:
:\re^{b\p^+ (\xip_2 )}:  &=&
:\re^{\left( a\p^+ (\xip_1 ) + b\p^+ (\xip_2 ) \right)}:  \nn  \\
 && \cdot   \mid 1- \re^{i( \xip_1 -\xip_2 )}
       \mid^{-ab {{\g^2}\over {4\pi}}}
  \re^{-{{i\g^2} \over 8} ab \e (\xip_1 -\xip_2) }
  \re^{iab {{\g^2}\over {16\pi}} (\xip_1 - \xip_2) } \nn\\
&& \label{OPEexp1}
\eqaend
and
\eqabegin
:\re^{ a\p^+ (\xim_1 )}:
:\re^{b\p^+ (\xim_2 )}:  &=&
:\re^{\left( a\p^+ (\xim_1 ) + b\p^+ (\xim_2 ) \right)}:  \nn  \\
 &&  \cdot  \mid 1- \re^{i( \xim_1 -\xim_2 )}
       \mid^{-ab {{\g^2}\over {4\pi}}}
  \re^{-{{i\g^2} \over 8} ab \e (\xim_1 -\xim_2) }
  \re^{iab {{\g^2}\over {16\pi}} (\xim_1 - \xim_2) } \nn\\
  && \label{OPEexp2}
\eqaend
so the product of two normal ordered exponentials equals the fully
normal ordered product up to a ``short-distance factor".
Multiplying these two expressions and noticing that, at equal times,
$\xip_1 - \xip_2 = - (\xim_1 - \xim_2 ) = \s_1 - \s_2 $, we see
that the phase factor cancels between the left and the right moving
sectors. The final result is therefore symmetrical, so that
for arbitrary weights $a$ and $b$, we deduce
\eqabegin
:\re^{a \p (\tau , \s_1 )}: :\re^{b \p (\tau , \s_2 )}: =
:\re^{b \p (\tau , \s_2 )}: :\re^{a \p (\tau , \s_1 )}:
    \qquad {\rm for} \;\;\;  \s_1 \neq \s_2 \period
\eqaend
This shows that two free-field exponentials are indeed
mutually local, although this property does not hold for the
left- and right-moving sectors separately because of the
extra phase factors.

It is also well known that the normal ordered
free field exponentials behave properly under conformal
transformations. With the normal ordered Virasoro generators
\eqabegin
L_m^{(\pm )} = \ft12 \sum_n :a_n^{(\pm )} a_{m-n}^{(\pm )}:
         + {{\sqrt{4\pi}}\over{\g}} m a^{(\pm)}_m \comma
\eqaend
one obtains
\eqabegin
[ L_m , L_n ]  = (m-n) L_{m+n} + {c \over {12}}
                 m^3 \d_{m+n,0} \comma
\eqaend
where the central charge is now given by (\ref{ccharge});
as we explained there,
the contribution from the renormalized coupling constant,
which appears already in the classical algebra (\ref{Viralg}),
is augmented by $\ft1{12}$ from the normal ordering of the oscillators
above.  A standard calculation shows that
\eqabegin
\Big[ L_m^\pm \, , \, :\re^{a \p (\ta , \s )}: \Big] =
\re^{im\xi^\pm} \left( - i\del_\pm +
   m (a - {{\g^2}\over {8\pi}} a^2 ) \right)
:\re^{a \p (\tau , \s )}: \period \label{Lexp1}
\eqaend

The customary Euclidean formulation of conformal field theory makes
use of operator product expansions. These are most conveniently derived
by the use of Wick's theorem. For convenience, we define the variables
\eqabegin
    z^\pm = \re^{i(\ta \pm \s )}
\eqaend
(or just $z,w,...$ if there is no danger of confusion). As is well
known, these variables can be analytically continued to
imaginary times by the replacement $\ta \rightarrow
-i\ta$ (in the appendix, we will provide further details
of the transcription between Minkowski and Euclidean conventions).
It is now elementary to show that
\eqabegin
   \del_+ \p (z) \del_+ \p (w) = : \del_+ \p (z) \del_+ \p (w) :
+ {{\g^2}\over {4\pi }} {{wz}\over {(z-w)^2}}\period\label{OPEpsi}
\eqaend
Here, the $i\e$-prescription introduced above is understood;
in the Euclidean formulation, it becomes unnecessary as the
singularity at coincident points is avoided by radial ordering.
Now using Wick's theorem to evaluate such products, we can derive
the important operator product expansion
\eqabegin
T_{++} (z )T_{++}(w )= {c\over 2} {{z^2 w^2}\over {(z - w)^4}}
 + {{2zw} \over{(z - w )^2}} T_{++} (w )
 + {{zw}  \over{(z - w )}} \del T_{++} (w )\comma \label{TT}
\eqaend
 where $\del$ now denotes the partial derivative with respect
$w^+$ (and not $\xip$); as is customary in conformal field
theory, we have dropped the non-singular terms on the right hand
side. From the formulas given in the appendix, one can easily verify
that the somewhat unusual factors of $zw$ disappear upon conversion
of these formulas into Euclidean language.
%%%%%%%%%%%%%%%%%%%%%%%%%%%%%%%%%%%%%%%%%%%%%%%%%%%
\Subsection{Quantum Definition of $\exp (\l \ve)$ }
%%%%%%%%%%%%%%%%%%%%%%%%%%%%%%%%%%%%%%%%%%%%%%%%%%%%%%
Our aim in this section is to define the quantized exponential of the
(interacting) Liouville field $\ve$, which will provide the
necessary ``gravitational dressing" of string vertex operators in a
non-trivial gravitational background. For lack of a better notation,
we will continue to designate the renormalized exponential operators
by $:\re^{\l \ve}:$, although the semicolons now mean something
different from and should not be confused with ordinary free field
normal ordering. The obvious point of departure for this construction
is the expansion (\ref{explphi1}) of the Liouville exponential in
terms of the free field $\p$. Since the quantization of free fields
is well understood (see foregoing section), one hopes that the
interacting theory can also be exactly quantized by exploiting
this relation. However, matters are not so easy, and although
considerable progress can be made, there remain some unresolved
problems, most notably the question of whether the expansion in
$\m^2$ makes sense at all. The main difficulty is the
complicated non-local and non-linear character of the relation between
the Liouville field and $\p$. So, for instance, it will become
apparent that free field normal ordering by itself is not sufficient
to achieve consistency with the two main requirements, which we expect
the exponential operators to satisfy. These are:
\medskip

\noindent
(i) The exponential operator $:\re^{\l \ve}:$ should transform
properly under the conformal group, i.e.
\eqabegin
\Big[ L_m^\pm \, , \, :\re^{\l \ve (\ta , \s )}: \Big] =
\re^{im\xi^\pm} \left( - i\del_\pm + m \D_\pm (\l ) \right)
:\re^{\l \ve (\tau , \s )}: \period
\eqaend
Here $\D_\pm (\l )$ denotes the renormalized (``dressed") conformal
dimension of the exponential operator which, in general, will be
different from the classical dimension as we already pointed out.
\medskip

\noindent
(ii) Two exponential operators at spacelike distances should commute,
i.e. they should be mutually local:
\eqabegin
\Big[ :\re^{\l \ve (\tau , \s )}: \, , \,
   :\re^{\n \ve (\tau , \s' )}: \Big] = 0  \qquad {\rm for} \;\;\;
\s \neq \s' \period \label{locality1}
\eqaend
\medskip

To evaluate condition (i), it is crucial that we use the free field
energy momentum tensor to define $L_m^{(\pm)}$ since there is no
way to make sense out of $T_{\pm \pm} (\ve )$ before the Liouville
field $\ve$ itself has been defined as a
quantum operator. It is an open question,
whether $T_{\pm \pm} (\ve )$ can be defined as a quantum operator
and whether this operator coincides with the free field
energy momentum tensor.

As we have shown in the preceding section, both requirements
are met for normal ordered exponentials of free fields, but they
are far from trivial to satisfy for the full Liouville operator
and will lead to stringent and unexpected constraints.
For instance, the obvious idea of defining $:\re^{\l \ve}:$ by simply
putting (free field) normal ordering symbols around the full sum
which defines $\re^{\l \ve}$ in terms of the free field at the
classical level, fails because it clashes
with the first condition (as will become clear in a moment).
The locality condition is not considered in the customary
Euclidean formulation. This omission could be justified if one
had an Osterwalder-Schrader type reconstruction theorem for Liouville
theory, which is, however, not available at this point. This is one
of our main reasons for sticking with a Minkowskian world-sheet:
the locality condition will lead to further restrictions on
the theory, which are ``invisible" in the Euclidean formulation.
These will be discussed in the following section.

One of the additional elements needed in the construction of the
full theory is a finite multiplicative (i.e. wave function)
renormalization of the free field, whose necessity was first pointed
in \cite{GN}. We thus rescale $\p$ according to
\eqabegin
\p \longrightarrow \eta \p \comma \label{rescale}
\eqaend
where the constant $\eta$ will be determined shortly.
The rescaling of $\p$ must be performed wherever $\p$ appears,
i.e. in particular inside the expression for $Y$. The definition of
$c(\a )$ (cf. (\ref{alpha}) must be modified accordingly, viz.
\eqabegin
\log \a = \ft12 \eta \g P \period
\eqaend
As another example, consider formula (\ref{OPEpsi}), which is replaced by
\eqabegin
   \del_+ \p (z) \del_+ \p (w) = : \del_+ \p (z) \del_+ \p (w) :
             + 2g{{wz}\over {(z-w)^2}} \comma
\eqaend
where, for later convenience (and in order to remain in unison
with the literature), we have introduced the constant $g$
\eqabegin
g \equiv {{\eta^2 \g^2} \over {8\pi }} = 2 \hbar \eta^2 \period
\eqaend
The renormalization implied by the rescaling (\ref{rescale}) will entail
a shift in the canonical conformal dimension of vertex
operators and is therefore at the origin of ``gravitational
dressing".

The necessity of introducing $\eta$ can be seen as follows.
The expansion (\ref{explphi1}) contains arbitrary powers of $Y$.
Clearly, a definite conformal dimension can be assigned to the infinite
sum if and only if $Y$ has conformal dimension zero, and the full
conformal weight is carried by the prefactor, which is local in
the free field $\p$. So, we first make $Y$ well defined, replacing
it by $:Y:$, where semicolons denote symmetric free field normal
ordering. As a consequence, the conformal weight of the full
expression is determined by the prefactor alone, which we now replace
by $:\re^{\l \eta \p}:$, taking into account the multiplicative
renormalization introduced above. To compute the conformal dimension
of $:Y:$, we first rewrite (\ref{Lexp1}) in the form
\eqabegin
\Big[ L_m^{(\pm )} \, , \, :\re^{\l \eta \p (\ta , \s )}: \Big] =
\re^{im\xi^\pm} \left( - i\del_\pm +
   m (\l \eta  - {{\g^2 \eta^2 }\over {8\pi}} \l^2 ) \right)
:\re^{\l \p (\tau , \s )}:\period \label{Lexp2}
\eqaend
In order to arrange for $:Y:$ to have conformal dimension zero,
we demand that
\eqabegin
\eta - {{\eta^2 \g^2}\over {8\pi}} =
       \eta - 2\hbar \eta^2 = \eta - g = 1 \period \label{etaeq}
\eqaend
Putting $\l =1$ in (\ref{Lexp2}) and assuming (\ref{etaeq}) to hold,
we get
\eqabegin
\Big[ L_m^{(+)} \, , \, :\re^{\eta \p (\ta , \s )}: \Big] =
-i\del_+ \left( \re^{im\xi^+}  :\re^{\eta \p (\tau , \s )}: \right)
\eqaend
(and the analogous result for $L_m^{(-)}$). Although the
exponentials of $\p^+$ and $\p^-$ appear with {\it different}
spatial arguments under the integral, this is already enough to
compute the commutator of $L_m^{(+)}$ with $:Y:$ because
$L_m^{(+)}$ commutes with the non-zero mode part of $\p^-$. As for the
zero modes, the momentum dependence of the integrand can be
ignored because $L_m^\pm$ is independent of $Q$.  Furthermore,
the dependence of the integrand on the center of mass coordinate
$Q$ is the same as in $:\re^{\eta \p (\tau , \s)}:$.
After a little algebra, we obtain
\eqabegin
[L_m^{(+)}  \, ,\, :Y(\ta , \s ):] &=&
   {{\m^2}\over 8} :\biggl[\,
  c(\a )^2 \int_0^{2\pi} d\s' \int_0^{2\pi} d\s'' \,
E_\a (\s - \s') E_{1\over \a} (\s- \s'' )   \nn    \\
 && \left( -i\deldel{\s'} \right) \left( \re^{im(\tau + \s' )}
                 \re^{\eta \p^+ (\ta + \s' )} \right)
\re^{\eta \p^- (\ta - \s'')} \,\biggr] : \period
\eqaend
This can now be integrated by parts. The boundary terms cancel, and
since the derivative on $E_\a$ is proportional to the $\d$-function
by (\ref{delE}), we can replace $\re^{im(\tau +\s')}$  in the integrand
by $\re^{im(\tau +\s)}$, which can be pulled out of the integral.
In this way, we arrive at the desired result
\eqabegin
[L_m^+ \, ,\, :Y(\ta , \s ):] = - i \re^{im\xip }
  {\del \over {\del \xip }} :Y(\ta , \s ): \period
\eqaend

Let us now return to equation (\ref{etaeq}); it is solved by
\eqabegin
\eta = {1\over {12}} \left( 25-d \pm
\sqrt{(25-d)(1-d)}  \right)\comma\label{eta}
\eqaend
where we have inserted the (renormalized) value for the
coupling constant in terms of $d$. This result shows not only that
$\eta$ depends on $d$ (the central charge of the matter system), but,
more significantly, that the solution imposes a rather stringent
constraint on $d$ itself, and thus on the dimension of the target
space in which the string moves. Since $\eta$ cannot be complex, we
must restrict $d$ to the ranges $d\leq 1$ or $d\geq 25$.
 Especially, for $d=1$ and $d=25$, one obtains $\eta =2$ ($g=1$)
and $\eta =0$ ($g=-1$), respectively; in the latter case, however,
the renormalized coupling constant diverges in such a way
that the product $\eta \g = 2\sqrt{2\pi} i$ stays finite.
In the remainder, we will be mostly interested in the borderline
case $d=1$\footnote{We
mention that for the unitary minimal models with
$d=1- {6\over {m(m+1)}}$, one obtains  two solutions $g_+ =
{m\over {m+1}}$ and $g_- = {{m+1}\over m} = g_+^{-1}$}.

The renormalized dimension of the exponential, which now comes
entirely from the factor $:\re^{\l \eta \p}:$, is given by
\eqabegin
\D (\l ) = \eta \l - {{\eta^2 \g^2} \over {8\pi}} \l^2
         = \eta \l - g \l^2 \period \label{deltal}
\eqaend
When coupling the exponential of the Liouville field to a string
vertex operator $\Phi$ of dimension $\D_0$, conformal invariance
requires that $\D (\l) + \D_0 = 1$, so that the total vertex operator
can be integrated over the world sheet. Classically, $\D (\l )= \l$,
so $\D_0 = 1- \l$. When $\D (\l)$ is deformed to the expression
above, $\D_0$ must be modified accordingly. Denoting the renormalized
(``gravitationally dressed") dimension of $\Phi$ by $\D$, a little
algebra shows that, with the above result, we must demand
\eqabegin
\D - \D_0 = 2\hbar \eta^2 \D_0 ( 1- \D_0 ) \label{delta}
\eqaend
in order to maintain total conformal dimension one. Formula (\ref{delta})
is nothing but the famous KPZ condition \cite{KPZ}.

In summary, requiring correct behavior of the exponential operator
with respect to conformal transformations, we have been led to the
result
\eqabegin
:\re^{\l \ve}: =: \re^{\l \eta \p}:
\sum_{n=0}^{\infty} {{(-1)^n}\over {n!}}
{{\G (2\l +n)}\over {\G (2\l )}} \left( :Y: \right)^n \label{explphi2}
\eqaend
with $\eta$ given by (\ref{eta}). The renormalized (``dressed")
dimension of this operator is given by (\ref{deltal}).
Let us stress once more that this
formula {\it defines} what we mean by the semicolons on
the left hand side of this equation. If instead, we had
defined the renormalized operator by free field normal ordering,
the result would have differed by certain short distance factors
(see also the discussion below), which would have led to additional
and anomalous contributions in the commutator of $L_m^\pm$ with the
exponential operator, and thus would have destroyed the nice
behavior of the operator under conformal transformations.

It must be emphasized at this point that the solution as written
down in (\ref{explphi2}) is not yet unique.
Since the Virasoro generators
do not depend on $Q$, and the desired commutation
relation with $L_m^{(\pm )}$ works order by order, we are
free to replace (\ref{explphi2}) by the modified expansion
\eqabegin
:\re^{\l \ve}: =: \re^{\l \eta \p}:
\sum_{n=0}^{\infty} {{(-1)^n}\over {n!}}
{{\G (2\l +n)}\over {\G (2\l )}}
  f_n (P)  \left( :Y: \right)^n \comma\label{explphi3}
\eqaend
where $f_n (P)$ are arbitrary functions of the center of mass
momentum $P$; the modified exponential operator behaves properly under
the conformal group. One can exploit (and eliminate) this
remaining freedom by imposing locality.

However, there is a price to pay. Even ignoring questions concerning
the convergence of the expansion, one may wonder whether the
individual terms in this expansion are actually well defined in
view of the appearance of arbitrary powers of $:Y:$ in it.
Naively and from one's experience with ordinary conformal field
theory, one would not anticipate difficulties, as $:Y:$ has
dimension zero by construction, and therefore the product of
two such operators should be non-singular at coincident points
if there are no negative dimension operators in the theory.
However, trouble is caused by the non-local
dependence of $:Y:$ on $\p^\pm$, which will lead to non-integrable
singularities inside the integrals defining $:Y:$. To make these
explicit, consider the product $:Y(\ta ,\s_1 ):: Y(\ta , \s_2 ):$.
Making use of (\ref{OPEexp1}) and (\ref{OPEexp2})
to rewrite this as a fully
normal ordered expression, we get additional short distance factors
inside the integral. Concentrating on
the ``critical region", where the integration
variables coincide, we can approximate the integral by
\eqabegin
&& :Y(\ta , \s_1 ): : Y(\ta , \s_2): \nn \\
&=& \int d\xi'_1 \int d\xi'_2 \int d\xi_1'' \int d\xi_2''  \,
\Big( \dots \Big)  {1 \over {(\xi_1' - \xi_2' - i\e )^{2g}}}
{1 \over {(\xi_1''- \xi_2''- i\e )^{2g}}} \comma
\eqaend
where we have only exhibited the singular factor, and the dots
stand for the harmless (and normal ordered) part of the integrand.
We repeat that we could have avoided this problem by defining the
exponential operator in terms of fully normal ordered products
right away, but at the expense of spoiling the conformal
properties, since $:\big(Y(\ta ,\s )\big)^n :$ is $not$ a conformal
field in the sense of (\ref{Lexp1}). Remarkably, the singularities
thus occur only in the integration variables, whereas the dependence
of the integrand on the ``external" variables
$\xip_1 = \ta + \s_1$ and $\xip_2 = \ta + \s_2$ is completely regular.
Performing two of the (indefinite) integrals, we end up with
\eqabegin
\int d\xi_1' \left( \xi_1' - \xi_2' - i\e \right)^{1-2g}
\int d\xi_1''\left( \xi_1''- \xi_2''- i\e \right)^{1-2g}
\Big( \dots  \Big)
\eqaend
These integrals exist for $g < 1$; at $g=1$, they are still
well defined because of the $i\e$ prescription.
This means that at low orders in the
expansion, the singularities are still integrable for a suitable
range of values of $g$; yet, at higher orders, the number of
poles at coinciding arguments will increase faster (namely
as $ n(n-1)/ 2$ for $(:Y:)^n$) than the number of integrations
(of which there are only $n$). Consequently, the singularities will
eventually become non-integrable for sufficiently large $n$,
no matter how we choose $g$. The problem disappears only at
special values of the parameter $\l$ (namely when $2\l$ is a negative
integer), for which the series terminates after a {\it finite}
number of terms.
%
%%%%%%%%%%%%%%%%%%%%%%%%%%%%%%%%%%%%%%
\Subsection{Locality Condition }
%%%%%%%%%%%%%%%%%%%%%%%%%%%%%%%%%%%%%%
Having constructed the quantum Liouville operator, we now proceed
to exploit the consequences of the locality requirement stated in
(\ref{locality1}). We will find two (mutually incompatible) ways to
satisfy it. The first severely
restricts the allowed values of $\l$, but works to all orders.
The second solution, proposed in \cite{OW}, is based on a
modification of the exponential operator (\ref{explphi2}), which exploits
(and eliminates) the remaining freedom of choosing the $P$ dependence
of the expansion coefficients, cf. (\ref{explphi3}). This proposal
has the advantage that it salvages locality for $arbitrary$ values
of $\l$, but has so far only been shown to work to cubic order
in $\m^2$ \cite{OW}\footnote{See, however, the footnote in the
introduction.}; in fact, we will present the proof to lowest
non-trivial order only. To simplify the notation, we will drop the
normal ordering symbols, and set $\ta =0 $ for convenience.

For ease of comparison with \cite{OW}, we shall
trade the field $Y$ for another one, called $S$, which differs in that
the zero mode dependence has been pulled out. So we write
\eqabegin
 \re^{\lambda\rphi} = \re^{\lambda\eta\psi}
              \sum_m C_m(\Pbar, \lambda) S^m \comma
\eqaend
where we only indicate the dependence on the center of mass momentum
and adopt the normalization
\eqabegin
C_0(\Pbar, \lambda) = 1 \period
\eqaend
For convenience, we have also defined
\eqabegin
\Qbar \equiv  \gamma\eta Q, \qquad  \Pbar \equiv {1\over 4}\gamma\eta P
\comma \label{QPbar}
\eqaend
in terms of which the canonical commutator reads
\eqabegin
\big[ \Qbar \, , \, \Pbar \big] = 2\pi ig \period
\eqaend
The quantity $S$ is related to $Y$ and defined by
\eqabegin
  S(0,\s ) & =& \int d\sigma' d\sigma''
      \re^{\Qbar/2} \re^{\Pbar \thsig}  \re^{{\Pbar \over \pi}
     (\sigma' -\sigma'')} \re^{\Qbar/2} \nn\\
  && \quad  \cdot  e^{\eta\psi^+_c(\sigma')}
              \re^{\eta\psi^+_a(\sigma')}
              \re^{\eta\psi^-_c(-\sigma'')}
              \re^{\eta\psi^-_a(-\sigma'')}\comma
\eqaend
where
\eqabegin
\t (\s ) \equiv \t (\s ; \s', \s'' ) \equiv
 \epsilon(\sigma -\sigma') -\epsilon(\sigma -\sigma'') \period
\eqaend
It should be clear that this is essentially the same expression
as the one that we derived for $Y$ in section 2.1. The integrand
has been written in a slightly different fashion by means of
the above definition (\ref{QPbar}) and the identity
\eqabegin
 E_\alpha(\sigma-\sigma')E_{1/\alpha}(\sigma-\sigma'')
 &=& \exp\left( \bar{P}\left( \epsilon(\sigma-\sigma') -
  \epsilon(\sigma-\sigma'')\right) \right)\period
\eqaend
Evidently, the operators $Y$ and $S$ differ only in their
dependence on the momentum $\Pbar$. The precise relation is
\eqabegin
 Y &=& {8\over \mu^2}\left[C(\Pbar+i\pi g)\right]^2 S \comma\nn\\
  C(\Pbar) &=& {1\over 2\sinh(\Pbar)}\comma
\eqaend
where the shift in the argument of $C(\Pbar )$ comes from pulling
the $\Pbar$-dependence out of the integral and the
normal ordering symbols. Expanding both sides and comparing
term by term, one arrives at the result
\eqabegin
 C_m(\Pbar, \lambda) &=& {(-1)^m \over m!}
  {\Gamma(2\lambda+m) \over \Gamma(2\lambda)}{\mu^2\over 8}
  \left[C(\Pbar +i\pi g)\right]^{2m} \period \label{Cmclass}
\eqaend
It will be important to note that this expression is invariant under
the shift
\eqabegin
 \Pbar & \longrightarrow & \Pbar + i\pi n\qquad n\in {\bf Z} \period
\eqaend

For convenience, we here repeat the exchange relation (\ref{exchange})
in a form suitable for the analysis of the locality condition.
Not forgetting to take into account the rescaling (\ref{rescale}),
 we have
\eqabegin
  \re^{a\eta\psi^+_a(\sigma_1)} \re^{b\eta\psi^+_c(\sigma_2)}
 & = & \mid 1-\re^{i\abs{ \sigma_1 -\sigma_2} }\mid^{-2abg} \nn\\
 & & \quad \cdot  \re^{-i\pi abg\epsilon(\sigma_1 -\sigma_2) }
  \re^{iabg(\sigma_1-\sigma_2)}  \nn\\
 && \quad \quad \cdot \re^{b\eta\psi^+_a(\s_2)}
\re^{a\eta\psi^+_a(\sigma_1)}  \comma \nn\\
    \re^{a\eta\psi^-_a(\s_1)} \re^{b\eta\psi^-_c(\sigma_2)}
 & = & \mid 1-\re^{i\abs{ \sigma_1 -\sigma_2} }\mid^{-2abg} \nn\\
 & &\quad  \cdot  \re^{i\pi abg\epsilon(\sigma_1 -\sigma_2) }
  \re^{-iabg(\sigma_1-\sigma_2)} \nn\\
 & & \quad\quad \cdot \re^{b\eta\psi^-_a(\s_2)}
\re^{a\eta\psi^-_a(\sigma_1)} \comma
\eqaend
where $\p_a^\pm$ and $\p_c^\pm$ were already defined in (\ref{psia}) and
(\ref{psic}). The corresponding formulas for the zero modes read
\eqabegin
 \re^{a\bar{Q}} f(\bar{P}) &=& f(\bar{P}+ia2\pi g) \re^{a\bar{Q}}
 \comma \nn\\
 f(\bar{P}) \re^{a\bar{Q}} &=& \re^{a\bar{Q}} f(\bar{P}-ia2\pi g)
 \period
\eqaend
This means
\eqabegin
 \re^{a \bar{Q}} f_a(\bar{P}) \re^{a \bar{Q}}
\re^{b \bar{Q}} f_b(\bar{P}) \re^{b \bar{Q}} &=&
\re^{b \bar{Q}} f_b(\bar{P}+ia4\pi g) \re^{b \bar{Q}}
 \re^{a \bar{Q}} f_a(\bar{P}-ib4\pi g) \re^{a \bar{Q}}\period
\eqaend

We are now ready to analyze the consequences of the locality condition.
For this purpose, we expand
\eqabegin
\Big[ :\re^{\l \ve (\tau , \s )}: \, , \,
   :\re^{\n \ve (\tau , \rho )}: \Big] = 0  \qquad {\rm for} \;\;\;
\s \neq \rho \comma
\eqaend
and require the result to vanish order by order. At lowest non-trivial
order, we obtain the following condition
(the $O(1)$ contribution is trivial), which must hold for all
values of the arguments $\s$ and $\rho$
\eqabegin
 0 &=&  \re^{\lambda\eta\psi(\sigma)} \re^{\nu\eta\psi(\rho)}
  C_1(\Pbar, \nu)S(\rho) \nn\\
 & & \quad -  \re^{\nu\eta\psi(\rho)}C_1(\Pbar,\nu)S(\rho)
        \re^{\lambda\eta\psi(\sigma)} \nn\\
 & & \quad \quad +  \re^{\lambda\eta\psi(\sigma)}
           C_1(\Pbar, \lambda)S(\sigma)
                 \re^{\nu\eta\psi(\rho)} \nn\\
 & & \quad \quad\quad -  \re^{\nu\eta\psi(\rho)}
          \re^{\lambda\eta\psi(\sigma)}
           C_1(\Pbar, \lambda)S(\sigma) \period
\eqaend
Since the non-zero mode contributions coincide in all four terms,
we must only look at the zero mode contributions. It is now
straightforward to show that this leads to the following
conditions on the coefficients:
\eqabegin
 0 &=& C_1(\Pbar, \nu) \re^{(\Pbar+i\pi g)\thrho } \nn\\
  & & \quad- C_1(\Pbar-i\pi 2\lambda g, \nu)
      \re^{(\Pbar+i\pi g(1-2\lambda))\thrho
     +i\pi 2\lambda g\thsig } \nn\\
  & & \quad\quad+ C_1(\Pbar-i\pi 2\nu g, \lambda)
      \re^{(\Pbar+i\pi g(1-2\nu))\thsig
     +i\pi 2\nu g\thrho } \nn\\
 & & \quad\quad\quad-C_1(\Pbar, \lambda) \re^{(\Pbar+i\pi g)\thsig }
 \period
\eqaend
This equation must be satisfied in all the regions of $(\sigma, \rho)$
space.  From its definition, $\thsig$ is easily seen to take the
values
\eqabegin
 \thsig &=& \bracebegin{llll} \ 0 & \qquad \mbox{if}\qquad
                         \sigma > \sigma'\ , \sigma''
                                 \qquad \mbox{or}\quad
                          \sigma',\ \sigma'' > \sigma \\
                          \ 2 & \qquad \mbox{if}\qquad
                        \sigma'' > \sigma > \sigma' \\
                           -2 & \qquad \mbox{if}\qquad
                        \sigma' > \sigma > \sigma''
             \braceend \nn
\eqaend
Therefore, $\thsig$ and $\thrho$ can take separately the values
 $0,\ \pm 2$, except for the combination $\thsig = \pm 2,\
\thrho = \mp 2$.  It is not difficult to see that many of the regions
give identical conditions and there are only three independent
equations (plus the corresponding ones in which $\lambda$ and $\nu$ are
interchanged):
\eqabegin
 0 &=& C_1(\Pbar, \nu) -C_1(\Pbar -i\pi 2g\lambda, \nu) \nn\\
   & & \quad + C_1(\Pbar-i\pi 2g\nu, \lambda) -C_1(\Pbar, \lambda)
      \label{loc1}\comma \\
 0 &=& C_1(\Pbar, \nu) \re^{2(\Pbar+i\pi g)}
      -C_1(\Pbar-i\pi 2g\lambda, \nu) \re^{2(\Pbar+i\pi g(1-2\lambda))}
   \nn\\
   & & \quad + C_1(\Pbar-i\pi 2g\nu, \lambda)  \re^{2\pi i 2\nu g}
 -C_1(\Pbar, \lambda)  \label{loc2} \comma \\
 0 &=& C_1(\Pbar, \nu)\re^{-2(\Pbar+i\pi g)}
      -C_1(\Pbar-i\pi 2g\lambda, \nu) \re^{-2(\Pbar+i\pi g(1-2\lambda))}
  \nn\\
  & & \quad + C_1(\Pbar-i\pi 2g\nu, \lambda) \re^{-2\pi i 2\nu g}
 -C_1(\Pbar, \lambda) \label{loc3}
\eqaend
By subtracting $(\ref{loc2})$ and $(\ref{loc3})$ from
$(\ref{loc1})$, we get
\eqabegin
 & & C_1(\Pbar, \nu)\left(1-\re^{2(\Pbar +i\pi g)}\right) \nn\\
 & & \quad -C_1(\Pbar-i\pi 2g\lambda, \nu)
     \left( 1- \re^{2(\Pbar+i\pi g(1-2\lambda))} \right) \nn\\
 & & \quad\quad +C_1(\Pbar-i\pi 2g\nu, \lambda)
     \left( 1- \re^{4\pi i\nu g} \right) = 0
\comma \label{loc4} \\
 & & C_1(\Pbar, \nu)\left(1- \re^{-2(\Pbar +i\pi g)}\right) \nn\\
 & & \quad -C_1(\Pbar-i\pi 2g\lambda, \nu)
     \left( 1- \re^{-2(\Pbar+i\pi g(1-2\lambda))} \right) \nn\\
 & & \quad\quad+C_1(\Pbar-i\pi 2g\nu, \lambda)
     \left( 1- \re^{-4\pi i\nu g} \right) = 0 \period \label{loc5}
\eqaend
Notice that in both of these equations a factor of the form
$ 1- \re^{\pm 4\pi i\nu g} $ appears in the last term.
Although the notations are somewhat different, effectively this factor
was implicitly assumed to be non-vanishing in the existing analysis
\cite{OW}.  As we shall  show, the case where this factor
vanishes will yield a new solution which is neither smoothly connected
nor relatively local to the solution obtained in \cite{OW}.
Thus we now solve (\ref{loc4}) and
(\ref{loc5}) for these two cases separately. Since our analysis will
involve the consideration of special discrete values of $\l$, we
emphasize that, except for $d=1,25$, the equation for $g$ admits
{\it two} solutions.
 \parmedskipn
Case 1: \quad Begin with the case where the factor above vanishes.
We then have
\eqabegin
  2\nu g &=& \mbox{integer} \comma
\eqaend
and from (\ref{loc4}) and (\ref{loc5}) we easily obtain
\eqabegin
 C_1(\Pbar-i\pi 2g\lambda, \nu) &=&
 { 1- \re^{2(\Pbar +i\pi g)} \over 1-\re^{2(\Pbar+i\pi g(1-2\lambda))}}
  C_1(\Pbar, \nu) \comma \\
C_1(\Pbar-i\pi 2g\lambda, \nu) &=&
 { 1- \re^{-2(\Pbar +i\pi g)} \over 1-\e^{-2(\Pbar+i\pi g(1-2\lambda))}}
  C_1(\Pbar, \nu) \period
\eqaend
{}From the compatibility of these equations, we get
\eqabegin
  2\lambda g &=& \mbox{integer} \comma\nn\\
 C_1(\Pbar-i\pi 2g\lambda, \nu) &=& C_1(\Pbar, \nu)\period
\eqaend
It is easy to see that the classical form of the coefficient
(\ref{Cmclass}) satisfies this requirement. \parmedskipn
Case 2:\quad Now consider the case where $2\nu g \ne \mbox{integer}$.  After
 some
calculation using (\ref{loc4}) and (\ref{loc5}), we obtain
\eqabegin
 C_1(\Pbar, \nu) &=& {a(\nu) \over
  \sinh(\Pbar+i\pi g)\sinh(\Pbar +i\pi g(1-2\nu)) }\comma \label{loc6}
\eqaend
where $a(\nu)$ is an unknown function only of $\nu$. It is
determined by substituting (\ref{loc6}) into (\ref{loc4}).
 The result is
\eqabegin
 a(\nu) &=& a \sin(2\pi\nu g) \qquad ( a= const.)\period
\eqaend
Thus  we finally obtain
\eqabegin
 C_1(\Pbar, \nu) &=& {a \sin(2\pi\nu g) \over
  \sinh(\Pbar+i\pi g)\sinh(\Pbar +i\pi g(1-2\nu)) }\period \label{loc7}
\eqaend
It can be checked by a rather tedious calculation that this form of
the coefficient does satisfy the locality equations in all the regions
without any additional conditions on the values of $\lambda$ and
$\nu$. Apart from an inessential normalization, this is
the form obtained in \cite{OW}. Clearly, this
formula is very suggestive of a hidden quantum group structure of
quantum Liouville theory, and this has led the authors of \cite{OW}
to conjecture a formula for the quantum deformed Liouville operator
to all orders\footnote{ The quantum group structure of Liouville theory
has also been extensively discussed in \cite{Gs}, although from
somewhat different point of view.}.
\par
%%%%%%%%%%%%%%%%%%%%%%%%%%%%%%%%%%%%%%%%%%%%%%%%%%%%%%%
Let us summarize the implications of the $\calO (\mu^2)$ analysis just
 performed :
First, if $2\nu g$ and $2\lambda g$ are {\it both} not integers, then
$\exp(\nu\varphi)$ and $\exp(\lambda\varphi)$ are mutually local with
the choice of the coefficients given by Otto and Weigt. On the other
 hand, if $2\nu g \in {\bf Z}$, then $\exp(\lambda\varphi)$ is
 mutually local with respect to $\exp(\nu\varphi)$ only if
$2\lambda g \in {\bf Z}$.  This latter case is precisely the one which
is relevant for the Liouville theory since the operator $\re^{\varphi}$
 appearing in the equation of motion must certainly exist.  One might
wonder if Case 1 is but a special case of Case 2.  This is not
so: In order for  the solution for Case 2 to correctly reproduce the
classical limit (\ie $g \rightarrow 0$ ),  the constant $a$ in
(\ref{loc7}) must be proportional to $1/\sin(\pi g)$, namely
\eqabegin
 C_1(\bar{P}, \nu) & \propto & {\sin(2\pi \nu g)/\sin(\pi g)
 \over \sinh (\Pbar+i\pi g) \sinh(\Pbar+i\pi g(1-2\nu))} \period
\eqaend
( Strictly speaking, the factor $\sin(\pi g)$ can be replaced by
any function going linearly to zero as $g\rightarrow 0$.)  However,
 if we now take the limit $\nu \rightarrow n/(2g)$ where $n$ is a
non-zero integer,  then the coefficient $C_1(\Pbar, \nu)$ above
vanishes.  Therefore the cases 1 and 2 above are not smoothly connected.
\par
Having seen that we must take $2g \in {\bf Z}$ for the Liouville theory
 to make sense, we will now show  that the locality condition is
satisfied to all orders in $\mu^2$ by taking the classical form of the
coefficients $C_m(\bar{P}, \lambda)$. \par
Let us first consider the effect of commuting
the free-field exponential $\exp(\lambda\eta\psi)$ through various
operators.  When commuted through $\bar{P}$, it produces the shift
$\bar{P} \longrightarrow  \bar{P}+i\pi2g\lambda =
\bar{P} + i\pi $ where $n$ is an integer. Since the coefficient
$C_m(\bar{P},\lambda)$ has the periodicity
\eqabegin
  C_m(\bar{P}+ i\pi n, \lambda) = C_m(\bar{P}, \lambda) \comma
\eqaend
it does not affect $C_m(\bar{P})$. Also the operator
\eqabegin
E_\alpha(\sigma-\sigma')E_{1/\alpha}(\sigma-\sigma'')
 = \exp\left( \bar{P}(\epsilon(\sigma-\sigma') -
  \epsilon(\sigma-\sigma'')\right)
\eqaend
is not affected,
since $\epsilon(\sigma-\sigma') - \epsilon(\sigma-\sigma'')$ is always
an even integer.

One more change $\exp(\lambda\eta\psi)$ produces when commuted with
the operator $S(\rho)$ is a phase of the type $\exp(i(\rho' -\rho'')a)$.
{}From the exchange formula, we can check that while the non-zero mode
exchange produces the phase $\exp(-i(\rho'-\rho'')2\lambda g)$, it is
precisely compensated by the opposite phase coming from the zero-mode
 exchange.  Thus we conclude that, for the case under consideration,
$\exp(\lambda\eta\psi)$ has no effect in the locality equation and
 hence can be ignored.
It is therefore sufficient to prove
\eqabegin
 & & \sum_{m,n}\left\{ C_m(\bar{P},\lambda)S^m(\sigma)
   C_n(\bar{P},\nu)S^n(\rho)\right. \nn\\
 & & \left. -C_n(\bar{P},\nu)S^n(\rho)
  C_m(\bar{P},\lambda)S^m(\sigma) \right\} =0 \period
\eqaend
Since $S^m(\sigma)$ contains the zero mode part $\exp(m\bar{Q})$, the
left hand side of the above equation becomes
\eqabegin
& & \sum_{m,n}\left\{ C_m(\bar{P},\lambda)C_n(\bar{P}+2i\pi gm,\nu)
 S^m(\sigma)S^n(\rho) \right. \nn\\
& & \left. -C_n(\bar{P},\nu)C_m(\bar{P}+2i\pi gn,\lambda)
 S^n(\rho)S^m(\sigma) \right\} \period
\eqaend
For each term in the sum, the exchange of $S^n(\rho)$ and $S^m(\sigma)$
produces the phase
\eqabegin
 S^n(\rho)S^m(\sigma) &=& S^m(\sigma)S^n(\rho)
  \re^{\left( 2\pi igm \sum\theta(\rho,j) -2\pi i gn
\sum\theta(\sigma,i)\right)} \period
\eqaend
But since each sum in the exponent is an even integer, the phase factor
is actually $1$ for $2g \in {\bf Z}$.  Also, in this case, we have
$C_n(\bar{P}+2\pi igm, \nu) = C_n(\bar{P})$ etc.. This means that
all the operators appearing in the locality equations commute and
therefore the locality requirement is satisfied to all orders
in $\mu^2$ with the classical expression for $C_m(\bar{P},\lambda)$.
%%%%%%%%%%%%%%%%%%%%%%%%%%%%%%%%%%%%%%%%%%%%%%%
\Subsection{Operator Equation of Motion}
%%%%%%%%%%%%%%%%%%%%%%%%%%%%%%%%%%%%%%%%%%%%%%%
The procedure by which we constructed the exponential operator
(\ref{explphi3}) relied heavily on the existence of the associated free
field $\p$, which itself was built out of the solution of the classical
Liouville equation in terms of the functions $A,B$. We can now ask
under what circumstances this equation can remain valid in the
quantized theory. The main question we have to address here is how
to define the quantum Liouville field itself (and not its exponential).
This is a rather subtle issue in view of the results of the preceding
section. If the proposal of \cite{OW} could be shown to work to all
orders, we could {\it define} the quantum field $\ve$ by taking the
derivative of $:\re^{\l \ve}:$ with respect to $\l$ and putting
$\l = 0$ afterwards. If, on the other hand, only the discrete and
mutually local set of operators with $2g\l \in {\bf Z}$ is available,
this idea does not work. In this case, we can define the Liouville
field by its normal ordered expansion, i.e.
\eqabegin
:\ve : \equiv \eta \p + \sum_{n=1}^\infty {{(-1)^n}\over n}
 \left( :Y: \right)^n  \period \label{ordphi}
\eqaend
Of course, we must keep in mind that this definition is afflicted with
the same problems that we encountered in defining the exponential
operator, especially since the expansion (\ref{ordphi}) contains
infinitely many terms (cf. the discussion at the end of section 3.2);
furthermore, with this definition the relation between $:\ve:$ and
the exponential operator defined before is somewhat obscure.
With this caveat, let us proceed nonetheless
and analyze the quantum Liouville equation. To do so, we need some
relations involving derivatives of $:Y:$. From (\ref{delE})
 and (\ref{Y}), we easily derive
\eqabegin
:\partial_+ Y(\ta , \s ) :\,  =  : \left( c(\a )
  \int_0^{2\pi} d\s''   \,
 E_{1\over \a} (\s - \s'' ) \re^{\p^+ (\tau + \s )}
\re^{\p^- (\ta - \s'')}  \right) : \comma
\eqaend
and
\eqabegin
:\partial_- Y(\ta , \s ) :\,  =  : \left( c(\a )
  \int_0^{2\pi} d\s'   \,
 E_\a  (\s - \s' ) \re^{\p^+ (\ta + \s' )}
\re^{\p^- (\tau - \s )} \right) : \period
\eqaend
(We put $\m^2 =8$ for simplicity.) For the double derivative, we obtain
\eqabegin
:\del_+ \del_- Y(\ta , \s ): \, =
:\re^{ \eta \p^+ (\xip )} \re^{\eta \p^- (\xim )}: =
:\re^{\eta \p (\ta , \s )}:\period
\eqaend
A separate calculation shows that
\eqabegin
\big[ :\del_\pm Y(\ta , \s ): \, , \, :Y(\ta , \rho ): \big] = 0
\period
\eqaend
We can therefore differentiate (\ref{ordphi}) term by term and move
the derivatives to the right; the calculation is then almost
identical with the classical case. Using the above relations,
we thus get
\eqabegin
\del_+ \del_- \ve  &=& 2\sum_{n\geq 1}(-1)^n \left( :Y: \right)^{n-1}
          : \re^{\eta \p} : +  \nn   \\
  &+& 2 \sum_{n\geq 1} (-1)^n (n-1) \left( :Y: \right)^{n-2}
  :\del_+ Y : :\del_- Y:\period \label{ddphi}
\eqaend
Now, it is a little exercise in re-normal ordering to show that
\eqabegin
:\partial_+ Y(\ta , \s ) :\, :\del_- Y(\ta , \r ): &=&
 :\left( c(\a )  \int_0^{2\pi} d\s''   \,
       E_{1\over \a} (\s - \s'' ) \re^{\p^+ (\ta + \s )}
               \re^{\p^- (\ta - \s'')}  \right) :   \nn   \\
    &\cdot&   : \left( c(\a )
           \int_0^{2\pi} d\s'   \,
          E_\a  (\r - \s' ) \re^{\p^+ (\ta + \s' )}
            \re^{\p^- (\ta - \r )} \right) :
\eqaend
\eqabegin
= (-1)^{2g}     :\Big(
   c(\a )^2 \int_0^{2\pi} d\s' \int_0^{2\pi} d\s'' \, &&
E_\a (\r - \s') E_{1\over \a} (\s- \s'' ) \re^{\p^+ (\ta + \s' )}
\re^{\p^- (\ta - \s'')}\Big):  \nn   \\
&&  :\re^{\eta \p^+ (\s )} \re^{\eta \p^- (\s' )}: \period
\eqaend
Putting $\s = \r$, we get
\eqabegin
:\del_+ Y(\ta ,\s ): \, : \del_- Y(\ta ,\s ) : = (-1)^{2g} :Y(\ta ,\s ):
:\re^{\eta \p (\ta ,\s )}: \period
\eqaend
Inserting this into (\ref{ddphi}), we see that for $g \in {\bf Z}$,
the right hand side of (\ref{ddphi}) can be simplified to
\eqabegin
  2\sum_{n\geq 1} (-1)^n n \left( :Y: \right)^{n-1} : \re^{\eta \p}:
 = -: \re^\ve : \comma
\eqaend
which is the desired result.
We conclude that under the assumptions made above
the quantum Liouville equation is satisfied only for integer
values of $g$, and, in particular, for $d=1$. For the
$d<1$ models, $g$ is not an integer, as is obvious from (\ref{etaeq}).
In the foregoing section, we demonstrated that locality holds to all
orders if $2g\l$ is an integer. For non-integer $g$, $\l$ is not
integer either; but then, $:\re^\ve:$ cannot be a local operator.
This seems to indicate that the quantum Liouville equation is
not consistent with locality for $d<1$! However, we repeat that
this conclusion is subject to the caveats mentioned above, and
therefore does not necessarily imply any inconsistency of the
$d<1$ models.
%%%%%%%%%%%%%%%%%%%%%%%%%%%%%%%%%%%%%%%%%%
%%%%%%%%%%%%%%%%%%%
%  lv3-4.tex
%%%%%%%%%%%%%%%%%%%%
\section{Ground Ring}
%%%%%%%%%%%%%%%%%%%%
As an application of the formalism developed in the preceding sections,
we shall make an attempt to construct the generators of
the so called \lq\lq ground ring" \cite{Wn:GR}, which was found to play
 an important role in characterizing the symmetry structure of the $d=1$
string theory. Our prime concern here is to see how its structure
changes  when the dependence on the cosmological constant is fully
taken into account in the operator formalism.
%%%%%%%%%%%%%%%%%%%%%%%%%%%%%%%%%%%%%%%%%%%
\Subsection{Euclidean Case with $\mu^2=0$ }
%%%%%%%%%%%%%%%%%%%%%%%%%%%%%%%%%%%%%%%%%%%
To begin with, let us briefly recall the Euclidean case with $\mu^2$
 set to zero, the case first discussed by Witten \cite{Wn:GR}. When
the Liouville field is regarded as a free field, one can
consider the ring of BRST invariant operators with vanishing ghost
number for the holomorphic and the anti-holomorphic sectors separately.
Each of them is hence termed  \lq\lq a chiral ground ring".
In the holomorphic sector, the chiral ring was shown to be  generated
by the following two operators, called $x$ and $y$ :
\eqabegin
 x &=& \left(cb+{i\over \rttwo}(\del X-i\del\phi) \right)
    \cdot {\rm e}^{i(X+i\phi)/\rttwo}\comma  \\
 y &=& \left(cb-{i\over \rttwo}(\del X+i\del\phi) \right)
    \cdot {\rm e}^{-i(X-i\phi)/\rttwo} \comma
\eqaend
where $b(z)$ and $c(z)$ are the ghost fields, while $\phi(z)$ and $X(z)$
are respectively the Liouville and the matter field with the free-field
 operator products
\eqabegin
 \phi(z)\phi(w) &\sim& X(z)X(w) \sim -\ln (z-w) \period
\eqaend
These operators correspond to special discrete physical states and
are invariant with respect to the BRST operator $Q_E$ given by
\eqabegin
 Q_E &=& \sum c_{-n}L^E _n -
\half \sum (m-n):c_{-m} c_{-n} b_{m+n}:_{inv}\comma
\eqaend
where the subscript $inv$ signifies $sl(2)$ invariant normal ordering.
The Virasoro generator $L_n^E$ is of the form
\eqabegin
 L^E_n &=& L^X_n +L^\phi_n \comma \\
 L^X_n &=& \half \sum : \alpha^X_{n-m} \alpha^X_m : \comma \\
 L^\phi_n &=& \half \sum :\alpha^\phi_{n-m} \alpha^\phi_m :
              +iq(n+1)\alpha^\phi_n \comma
\eqaend
where the background charge $q$ takes the value $\rttwo$.
The anti-holomorphic chiral ground ring is generated by similar
operators called $\bar{x}$ and $\bar{y}$.  \par
In the following, we shall concentrate on the generator $x$ and
for ease of notation  express it as
\eqabegin
 x &=&  j\calx \Phizero +\del\calx \Phizero -\calx \del\Phizero \comma
  \label{xeuc}\\
j &\equiv& cb \comma\\
\calx & \equiv & \re^{ i {1 \over \sqrt{2}} X }\comma  \\
\Phizero & \equiv & \re^{ -{1 \over \sqrt{2}} \phi }\period
\eqaend
The subscript \lq\lq 0" emphasizes that the Liouville field is treated
as a free field and the usual normal ordering for composite operators is
understood.
%%%%%%%%%%%%%%%%%%%%%%%%%%%%%%%%%%%%%%%
\Subsection{Minkowski Case with $\mu^2=0$ }
%%%%%%%%%%%%%%%%%%%%%%%%%%%%%%%%%%%%%%%%%%%
Before tackling the fully interacting case, we need
to clarify in some detail how the free field case should be treated
in Minkowski formulation. Many of the calculations to be perfomed in this
subsection will be utilized in the interacting case.  \par
To facilitate the comparison with the Euclidean case, we shall use
the canonically normalized Liouville field $\phi(\xplus)$ which,
in terms of the field $\psi(\xplus)$, is given by
\eqabegin
 \phi(\xplus) &=& \overrttwo \psi(\xplus) \period
\eqaend
( To be precise, $\phi(\xplus)$ is defined to be that part of
$\phi(\xplus,\xminus)$ which is independent of $\xminus$. That is, it
includes the full $Q$-zero mode of $\phi(\xplus,\xminus)$. )
Then the energy-momentum tensor for the Liouville sector is given by
\eqabegin
 T^\phi_M (\xplus) &=& \half (\delplus\phi)^2 -q\delplus^2\phi \comma
\eqaend
and the corresponding Virasoro generator is of the form
(still with $q= \sqrt{2}$)
\eqabegin
L^{M,\phi}_n &=& \half \sum :\alpha^\phi_{n-m} \alpha^\phi_m :
              +iqn\alpha^\phi_n \period
\eqaend
Notice that the term containing the background charge takes a slightly
 different form compared with the Euclidean case.  It is
straightforward to show that the exponential operator $\Phizero$
 is primary with respect to this Virasoro generator {\it provided} that
it is defined with symmetric normal ordering (see the appendix.)
  Specifically,
\eqabegin
\left[ L^{M,\phi}_n, \Phizero(x_+)\right]
     &=& z^n \left( \oneoveri \delplus + \left(-{5\over 4}\right)n\right)
        \Phizero(x_+) \period
\eqaend
It is easy to check that $L^{M,\phi}_n$'s satisfy the Virasoro algebra
of the form
\eqabegin
 \left[L^{M,\phi}_m, L^{M,\phi}_n \right]
   &=& (m-n)L^{M,\phi}_{m+n} + \delta_{m+n,0}\left(
 {1+12q^2 \over 12}(m^3-m) + q^2m \right) \period
\eqaend
In order to construct the correct nilpotent BRST operator, we need
 Virasoro generators which satisfy the above algebra with the
central term proportional to $m^3-m$.  As is well-known, such a
standard form can be  obtained by a shift
\eqabegin
 \tilde{L}^{M,\phi}_n &=& L^{M,\phi}_n + \half q^2 \delta_{n,0}
 = L^{M,\phi}_n + \delta_{n,0} \period
\eqaend
If we denote the total Virasoro operator (including the matter part
 $L^{M,X}_n$) by $L^M_n$, the BRST operator takes the form
\eqabegin
 Q_M &=& \sum c_{-n}\left(L^M_n+\delta_{n,0}\right)
  -\half \sum (m-n)
 :c_{-m}c_{-n}b_{m+n}:_{inv} \comma \label{brsmin1}
\eqaend
\par
%%%%%%%%%%%%%%%%%%%%%%%%%%%%%%%%%%%%%
We are now ready to display the generator $x_M$ , corresponding to
the operator $x$, for the  Minkowski case.  It takes the form
\eqabegin
 x_M &=& \left(j-{3\over 2}\right)\calx \Phizero + \overi \delplus
 \calx\Phizero -\overi \calx \delplus \Phizero \comma
\eqaend
where the ghost current $j$ is normal-ordered with respect
to the $sl(2)$ invariant vacuum as in the Euclidean case,
 and the operators $\Phizero$ and
$\calx$ are defined with the symmetric normal ordering.  The shift of
$j$ by the amount $3/2$ compared with (\ref{xeuc}) can be
understood as follows: The BRST operator $Q_M$
 given in (\ref{brsmin1}) actually takes a better form
if we make a change of normal-ordering
from the $sl(2)$ invariant one to the \lq\lq physical" one
defined by the rule
\eqabegin
 b_n,\ c_n & & \mbox{annihiliation for $n \ge 1$} \comma\\
 \half ( c_0 b_0 -b_0 c_0 ) & & \mbox{ for zero mode} \period
\eqaend
$Q_M$ then becomes
\eqabegin
 Q_M &=& \sum c_{-n}L^M_n
  -\half \sum (m-n)
 :c_{-m}c_{-n}b_{m+n}:_{phys} \period \label{brsmin2}
\eqaend
It is easily checked that under this normal ordering $j$ becomes
anti-hermitian and is related to the previous definition by
\eqabegin
 j^{inv}-{3\over 2} &=& j^{phys} \period
\eqaend
\par
BRST invariance of $x_M$ given above  can be demonstrated as follows.
First we write the commutator $\left[Q_M, x_M\right]$ as
\eqabegin
 \left[ Q_M, x_M \right] &=& A_1 +A_2 +A_3 -{3\over 2}A_4 \comma
\eqaend
where
\eqabegin
 A_1 &=& \lim_{y^+ \rightarrow x^+} \left[ Q_M, j(x^+)
                  \calx\Phizero(y^+) \right] \comma\\
 A_2 &=& \left[ Q_M, -\calx\oneoveri\delplus \Phizero \right]\comma \\
 A_3 &=& \left[ Q_M, \oneoveri \delplus \calx \Phizero \right]\comma \\
 A_4 &=& \left[ Q_M, \calx \Phizero \right] \period
\eqaend
$A_2$, $A_3$ and $A_4$ are easy to evaluate using the fact that $\calx$
 and $\Phizero$ are primary with respect to the Virasoro operators
 inside $Q_M$.  The result is
\eqabegin
 A_2 &=& c\delplus\left(\calx \delplus \Phizero \right)
         -{5\over 4}\delplus^2 c \calx \Phizero \comma\\
 A_3 &=& -c \delplus \left(\delplus \calx\Phizero \right)
          -{1\over 4}\delplus^2 c \calx\Phizero  \comma\\
 -{3\over 2} A_4 &=& -{3\over 2} c \oneoveri\delplus\left(
            \calx\Phizero \right)
               +{3\over 2}\oneoveri\delplus c \calx\Phizero \period
\eqaend
Therefore,
\eqabegin
  A_2 + A_3 -{3\over 2} A_4 &=& -c\left( \delplus^2\calx \Phizero
                    -\calx \delplus^2 \Phizero \right)
                   -{3\over2}\delplus^2 c \calx\Phizero \nn\\
 & &\quad -{3\over 2} c \oneoveri\delplus\left(
            \calx\Phizero \right)
               +{3\over 2}\oneoveri\delplus c \calx\Phizero \period
  \label{A234}
\eqaend
\par
To compute $A_1$, we will need the following formulae which can be
obtained by straightforward calculations:
\eqabegin
 \left[ Q_M, j(x^+)\right] &=& -c\left(T_M(x^+)+1\right) -\oneoveri
     c\delplus c b +{3\over 2}\delplus^2 c
                   -{3\over 2}\oneoveri\delplus c \comma \\
 \left[ Q_M, \calx\Phizero(y^+)\right]
           &=& c\oneoveri \delplus \left( \calx\Phizero \right)
               -\oneoveri \delplus c \calx\Phizero \comma\\
 T^x_M(x^+)\calx (y^+) &=& \half\left({1\over \sqrt{2}}{w\over z-w}
                        + {1\over 2\sqrt{2}}\right)^2\calx
                     +\left( {w\over z-w}+\half\right)
                    \oneoveri\delplus\calx \comma\\
          & & \quad + {1\over \sqrt{2}}\oneoveri \delplus^2 X\calx
           + :T^x_M(x^+)\calx : \comma \\
 T^\phi_M (x^+)\Phizero(y^+) &=&
            -{5\over 4}{zw \over (z-w)^2}\Phizero
        + \left({w\over z-w}+\half\right) \oneoveri \delplus \Phizero
           \nn\\
        & & \quad + \left( {1\over \sqrt{2}}\delplus^2\phi
            -{1\over 16}\right) \Phizero
          + :T^\phi_M(x^+)\Phizero : \comma\\
 j(x^+)c(y^+) &=& c(x^+)\left({w\over z-w}+2\right)+:j(\xplus)c(\yplus):
          \comma\\
 j(x^+)\oneoveri \delplus c(y^+) &=&
         c(\xplus) \left(\left({w\over z-w}\right)^2 +
    {w\over z-w} -1\right)
         -\oneoveri c\delplus c b \period
\eqaend
Using these formulae, it is easy to verify that all singular terms
cancel, and we are left with
\eqabegin
 A_1 &=&  -c\left(\half\left(\delplus X \right)^2
            -{i\over \sqrt{2}}\delplus^2 X
            +\half\left(\delplus \phi \right)^2
            -{1\over \sqrt{2}}\delplus^2 \phi\right) \calx\Phizero \nn\\
    & & \quad + {3\over 2}c\oneoveri \delplus \left(\calx\Phizero \right)
       +{3\over 2} \delplus^2c \calx\Phizero
     -{3\over 2}\oneoveri \delplus c
        \calx\Phizero \nn\\
       &=& c\left(\delplus^2\calx\Phizero
        -\calx \delplus^2\Phizero \right) \nn\\
    & &\quad  + {3\over 2}c\oneoveri \delplus \left(\calx\Phizero \right)
       +{3\over 2} \delplus^2c \calx\Phizero
        -{3\over 2}\oneoveri \delplus c
        \calx\Phizero \period \label{A1}
\eqaend
{}From  (\ref{A234}) and (\ref{A1}) it is evident that
the sum $A_1+A_2+A_3-{3\over 2}A_4$ indeed vanishes.
%%%%%%%%%%%%%%%%%%%%%%%%%%%%%%%%%%%%%%%%%%%%%%
%%%%%%%%%%%%%%%%%%%%%
\Subsection{Minkowski Case with $\mu^2\ne 0$ }
%%%%%%%%%%%%%%%%%%%%%
We now describe an attempt to construct the generators of the ground
ring for the interacting case.  As we switch on the cosmological term,
left and right moving modes are coupled and we must necessarily
consider the full ground ring.  In the free field case its generators
are obtained as products of chiral generators, such as
$x_M\bar{x}_M,\ x_M\bar{y}_M$, etc..
Since the interacting generators must reduce to these forms
in the limit $\mu^2 \rightarrow 0$, it is reasonable to take
our candidates to be obtained from these products by appropriate
 replacements of free fields by corresponding interacting fields.
More specifically, we shall consider in the following the operator
$a_1$ which is obtained from $x_M\bar{x}_M$ by the replacements
\eqabegin
 \calx(\xplus)\bar{\calx}(\xminus) & \longrightarrow &
 \calx (\xplus, \xminus)\comma \\
 \delplus\calx(\xplus) \bar{\calx}(\xminus) &\longrightarrow &
\delplus \calx(\xplus,\xminus) \comma\\
 \delplus\Phizero(\xplus) \del_- \bar{\Phizero}(\xminus)
 &\longrightarrow & \delplus\del_-\Phi(\xplus,\xminus)
 \comma  \qquad  \mbox{etc.}
\eqaend
Since the commutation relations with the right and left BRST charges can
be considered separately, we shall concentrate on the action of the
left charge.  Then we can effectively work with only the $x_M$
factor of $a_1$ ( with interacting fields ) and at the end of
the calculation we can multiply by $\bar{x}_M$ and make the replacements
above.  \parsmallskip
%%%%
We note the remarkable fact that the weight in the exponent of
$\Phi$ is precisely such that the expansion in powers of
$\mu^2$ breaks off after the first non-trivial order; namely,
we have exactly $\Phi = \Phizero (1+Y)$.  This allows us
to perform the calculations without any approximation.
 \parsmallskip
%%%%%%%%%%%%%%%%%%%%%%%%%%%%%%%%%%%%%%%%%
%
Let us now calculate the commutator $\left[Q_M, x_M\right]$.
All the calculations which make use of the {\it conformal properties}
 of the fields $\calx$ and $\Phi$ proceed exactly as in the free case.
 The only place we must not use the free-field expressions is the
computation of
$T^\phi(\xplus)\Phi(\yplus)$ and $\delplus^2\Phi$.  The result
 using the conformal poperties is
\eqabegin
 \left[Q_M, x_M\right] &=&
  -c\calx(x) \left\{ T^\phi_M(\xplus)\Phi(y)
     +\left( {5\over4} \left({w\over z-w}\right)^2
      +{5\over 4}{w\over z-w} + {1\over 16} \right)\Phi \right. \nn\\
  & & \left.  -\delplus^2\Phi -\left({w\over z-w}+\half\right)
    \oneoveri\delplus\Phi \right\} \period\label{Qx}
\eqaend
This is precisely the combination which vanished in the free field
calculation.

To simplify this expression it is useful to factorize $\Phi$ as
\eqabegin
 \Phi &=& \Phizero \left(1+Y\right) \comma
\eqaend
where $\Phizero$ is as defined for free theory and symmetric normal
ordering for $\Phizero$ and $Y$ is implicit.  Now we shall prove the
simple yet non-trivial result
\eqabegin
 \delplus\Phi &=& \delplus\Phizero (1+Y) \period \label{DPhi}
\eqaend
To prove this, we must show that $\Phizero \delplus Y$ vanishes.  First
 from the definition of $Y$ we easily get
\eqabegin
\delplus Y (x) &=& {\mu^2 \over 8}\int d\sigma''  \re^{\bar{Q}/2}
            C(\bar{P})\re^{-\bar{P}\epsilon(\sigma-\sigma'')}
           \re^{{\bar{P}\over \pi} \oneoveri\ln z}
           \re^{{\bar{P}\over \pi} \oneoveri\ln \bar{z}'' }
           \re^{\bar{Q}/2}\nn\\
    & &  \times \re^{\sqrt{2}\phi^+_c(z)}\re^{\sqrt{2}\phi^+_a(z)}
        \re^{\sqrt{2}\phi^-_c(\bar{z}'')}\re^{\sqrt{2}\phi^-_a(\bar{z}'')}
            \period
\eqaend
A better representation is obtained if we move the second
 $\re^{\bar{Q}/2}$ factor through the parts involving $\sigma''$.
Then we can express $\delplus Y(x)$ in the form
\eqabegin
 \delplus Y(x) &=& :\re^{\sqrt{2}\phi^+(\xplus)}: Y^{(-)} \comma\\
\eqaend
where
\eqabegin
  Y^{(-)}  &=&
{\mu^2 \over 8}\int d\sigma''C(\bar{P}-i\pi)
         \re^{-(\bar{P}-i\pi)\epsilon(\sigma-\sigma'')}
        \re^{ ({\bar{P}\over \pi}-i) \oneoveri\ln \bar{z}'' } \nn\\
       & & \times \re^{\sqrt{2}\phi^-_c(\bar{z}'')}
          \re^{\sqrt{2}\phi^-_a(\bar{z}'')}     \period
\eqaend
With this expression, we can  compute the operator product between
$\Phizero $ and $\delplus Y$  at different arguments.  We get
\eqabegin
 \Phizero (\xplus)\delplus Y(y) &=& :\re^{-{1\over \sqrt{2}}\phi^+(\xplus)
            -{1\over \sqrt{2}}\phi^-(x^-)}: \nn\\
       & &  \cdot :\re^{\sqrt{2}\phi^+(\yplus)}: Y^{(-)}(y) \nn\\
      &=& \re^{-\half\ln(w/z)} \left(1-{w\over z}\right) \nn\\
 & & \cdot:\re^{-{1\over \sqrt{2}}\phi^+(\xplus)+\sqrt{2}\phi^+(\yplus)
           -{1\over \sqrt{2}}\phi^-(x^-)}: Y^{(-)}(y) \period
\eqaend
This is seen to vanish in  the limit  $\yplus \rightarrow \xplus$ and
we have the announced result (\ref{DPhi}).\par
With the use of various formulae developed for the free field case, we then
get, after some calculations
\eqabegin
 \delplus^2\Phi
    &=& :T^\phi\Phizero:(1+Y)
      +:{1\over \sqrt{2}}\delplus^2\phi\Phizero :(1+Y)
        +\delplus\Phizero \delplus Y \period \label{DDPhi}
\eqaend
Now look at $T^\phi_M(\xplus)\Phi(y)$ in $[Q_M, x_M]$.
 We first re-normal order $T^\phi_M$ and $\Phizero$ and leave the factor
 $1+Y$ untouched.  The result is
\eqabegin
 T^\phi_M(\xplus)\Phi(y) &=&
     -\left( {5\over4} \left({w\over z-w}\right)^2
      +{5\over 4}{w\over z-w} + {1\over 16} \right)\Phi(y) \label{TPhi}\\
   & & +\left({w\over z-w}+\half\right) \oneoveri\delplus
        \Phizero (1+Y) \nn\\
   & & + :{1\over \sqrt{2}}\delplus^2\phi(y)\Phizero(y):(1+Y)
       +:T^\phi(\xplus)\Phizero(y): (1+Y) \period \nn
\eqaend
Substituting (\ref{DDPhi}) and (\ref{TPhi}) into (\ref{Qx})
 we get a {\it non-vanishing} result of order $\mu^2$:
\eqabegin
 \left[ Q_M, x_M \right] &=& c\calx \delplus \Phizero\delplus Y\period
\label{Qx2}
\eqaend
%%%%%%%%%%%%%%%%%%%%%%%%%%%%%%%%%%
\parsmallskip
%%%%%%%%%%%%%%%%%%%%%%%%%%%%%%%%%%%
%Check of BRST invariance of the Result
%%%%%%%%%%%%%%%%%%%%%%%%%%%%%%%%%%%
%
Since this result is not as expected,  let us make a few checks before
 contemplating upon its implications.
One check of the correctness of the calculation above is to show that
the right hand side of (\ref{Qx2})
\eqabegin
 A &\equiv & c\calx \delplus \Phizero\delplus Y \label{QxA}
\eqaend
 is $Q_M$ closed.
We need the following formulae:
\eqabegin
 \left\{ Q_M, c(\xplus) \right\} &=& c\oneoveri \delplus c(\xplus) \\
 \left[ Q_M, \Phizero\right] &=& c\oneoveri \delplus \Phizero
          -{5\over 4}\oneoveri \delplus c \Phizero \\
 \left[ Q_M, \delplus\Phizero\right] &=& \delplus c \oneoveri \delplus
       \Phizero + c\oneoveri \delplus^2 \Phizero \nn\\
     & & -{5\over 4}\oneoveri \delplus^2 c \Phizero
       -{5\over 4}\oneoveri \delplus c \delplus \Phizero \\
 \left[ Q_M, Y\right] &=& c\oneoveri \delplus Y \\
 \left[ Q_M, \delplus Y\right] &=& \delplus c \oneoveri \delplus Y
         + c\oneoveri \delplus^2 Y  \\
 \left[ Q_M, \calx\right] &=& c\oneoveri \delplus \calx
      + {1\over 4} \oneoveri \delplus c \calx
\eqaend
Using these expressions and $\Phizero \delplus Y =0$, we get
\eqabegin
 \left\{ Q_M, A \right\} &=&  c\oneoveri \delplus c \calx\delplus
       \Phizero \delplus Y \nn\\
  & & \quad -\left\{ {1\over 4} c\oneoveri \delplus c \calx\delplus
          \Phizero \delplus Y  \right.\nn\\
  & & \quad\quad -{1\over 4} c\oneoveri \delplus c \calx\delplus
          \Phizero \delplus Y \nn\\
  & & \quad\quad\quad \left. +c\oneoveri \delplus c \calx\delplus
       \Phizero \delplus Y \right\} \nn\\
  &= & 0 \period
\eqaend
Thus indeed (\ref{QxA}) is BRST closed. \parsmallskip
%%%%%%%%%%%%%%%%%%%%%%%%%%%%%%%%%%%%%%%
We now provide a further argument of more general nature which supports
 the result above.  This argument will lead to the conclusion that as
 long as we do not change the ghost structure of the operator $x_M$
it is not possible to construct a BRST invariant. \par
 It is well-known that the total zero-level Virasoro generator
$L^{tot}_0$, including the ghost part, can be written as $L^{tot}_0
 = \left\{ b_0, Q_M \right\}$.  On the other hand we have, for any
operator $\calO$ with a definite {\it global} dimension,
$ \left[ \Lzerotot, \calO\right] = \overi \delplus \calO $.
Combining these relations, we get
\eqabegin
 \left[\Lzerotot,\calO\right] &=& \overi\delplus\calO \nn\\
   &=& \left[\left\{ b_0, Q\right\},\calO \right] \nn\\
   &=& \left\{ b_0, \left[Q, \calO\right]\right\}
   + \left\{ Q, \left[b_0, \calO\right]\right\} \period
\eqaend
Thus if $\calO$ is $Q_M$-closed, then
\eqabegin
 \overi \delplus \calO &=& \left\{ Q, \left[b_0, \calO\right]\right\}
 \period
\eqaend
This equation states that  $\delplus\calO$ is necessarily BRST exact and
 is determined {\it solely} by the part of $\calO$ which contains the
 ghost $c$.\parsmallskip
%%%%%%%%%%%%%%%%%%%%%%%%%%%%%%%%%%%%%%%%%%%%%%%%
 Let us apply this logic to see if we can construct a BRST invariant
operator of the form
\eqabegin
 \calO &=& cb \calx \Phi + \mbox{terms not containing $c$}\period
\eqaend
For this class of operators, we have
\eqabegin
 \left[ b_0, \calO\right] &=& \left[b_0, cb\calx\Phi\right]
 = b\calx\Phi \period
\eqaend
Therefore
\eqabegin
 \overi\delplus\calO &=& \left\{ Q, b\calx\Phi\right\} \nn\\
  &=& \left\{Q, b\right\}\calx\Phi- b\left[Q, \calx\right]\Phi
 -b\calx\left[Q, \Phi\right] \period
\eqaend
Using the various formulae developed previously for the calculation
of $x_M$, we can compute the right hand side.  The result is
\eqabegin
 \delplus\calO  &=& \delplus \left( j\calx\Phi +\overi\delplus
\calx\Phi -\overi\calx\delplus\Phi -{3\over 2}\calx\Phi \right) \nn\\
 & & \quad + \overi \calx\delplus\Phi_0 \delplus Y \period
\eqaend
( One can check explicitly that $\left[ Q, \delplus\calO \right] =0$.)
 As expected, the expression in the parenthesis is precisely our
candidate $x_M$, but the additional term cannot be written as a total
derivative.  The closest we can get is
\eqabegin
 \overi \calx\delplus\Phi_0 \delplus Y  &=& \delplus \left(
 \overi \int_a^{\xplus} d\yplus \calx\delplus\Phi_0 \delplus Y(y)
   \right)\period
\eqaend
But the operator in parenthesis is not BRST invariant due to the
lower limit of the integration.   \parsmallskip
Thus we have shown that as long as we keep intact the structure
 of the term  involving the $c$-ghost, we cannot construct a BRST
invariant in terms of the the fully interacting field in analogy to
the free field case.  The situation does not
 improve even when we take into account the right-moving sector.
\parsmallskip
This result would mean that the structure of the ground ring remains
identical to that of the theory without interactions. Of course,
one might question our use of the free field form of the
BRST operator in reaching this conclusion. However, the only part
of this operator that could be plausibly affected and altered by the
interactions is the term with the Liouville energy momentum tensor.
We are hesitant about this possibility as it is difficult
to see how the construction can be modified without upsetting all
our results up to this point. After all, it is the free field
form of the Liouville energy momentum tensor that we have
been using in our construction of the exponential operator in
section 3.2 (since we did not even know how to {\it define} the energy
momentum tensor otherwise!). Despite the fact that we have
been using the free field form of the BRST operator, our conclusion
is not as trivial as it may seem. To underline this point, we note
that, as far as the conformal properties are concerned,
we do have  operators such as $\re^{\varphi}$ which exhibit the same
conformal properties as the corresponding free-field operator. It should
also be noted that, if we express the generators of the ring in terms
of the fully interacting fields, their forms certainly change into
complicated expressions involving $\mu^2$.  Since the result of the matrix
model, which incorporates the full interaction, indicates the existence
of the ground ring of the free-field structure, this interpretation
seems to be the correct one. The point is that the free-field is not the
 Liouville field, but it is a complicated combination of the
Liouville field.
%%%%%%%%%%%%%%%%%%%%%%%%%%%%%%%%%%%%%%%%%%%%%%%%%
%%%%%%%%%%%%%%%%%%%%%%%%%%%
% lv3-5.tex
%%%%%%%%%%%%%%%%%%%%%%%%%%%%%%%
\section{Discussion}
%%%%%%%%%%%%%%%%%%%%%%%%%%%%%%
Clearly, difficult problems remain.
The rigorous construction of the exponential
Liouville operator still has not been accomplished. Apart from the
technical difficulties discussed at the end of section 3.2, it is
far from clear whether a perturbative expansion in the cosmological
constant $\mu^2$ makes sense at all. As long as these problems are
not solved, there is little point in addressing other issues,
such as the question of whether the Liouville energy momentum tensor
can be consistently defined in terms of the interacting Liouville
field and shown to coincide with the free field energy
momentum tensor, or whether the transformation between the Liouville
field and the free field $\p$ is canonical also at the quantum level.
The hidden quantum group structure revealed in the construction of
\cite{OW} is certainly intriguing, but the existence of two mutually
exclusive solutions to the locality condition remains an
unsatisfactory feature. Perhaps the fact that there exist {\it two}
solutions for the constant $g$ (except for $d=1$ and $d=25$) plays a
role in resolving this issue. In any case, we find it remarkable that
the weights which appear in the ground ring operators are precisely
in agreement with our discretization condition (3.3.25) and such
that the expansion (3.2.15) has only finitely many terms.

A possible way out may be to shelve these questions for the time
being and to pursue the study of the operator formalism for Liouville
theory along the lines advocated by Gervais and collaborators
\cite{Gs},\cite{GsRev}. Rather than insisting on a rigorous
construction of the Liouville exponential operator, these authors
emphasize the importance of the hidden quantum group structure.
This point of view is supported by the fact that many results
can be deduced by requiring covariance with respect
to the hidden $SL(2)_q$, i.e. without invoking the explicit form of
the Liouville operator, and are found to agree with the results
obtained by other methods. A detailed comparison between this
approach and the results obtained in this paper certainly merits
further investigation.

%%%%%%%%%%%%%%%%%%%%%%%%%%%%%%%%%%%%%%%%%%
\vspace{1.5cm}\parn
{\Large\bf Acknowledgment} \parbigskipn
We would like to acknowledge stimulating and helpful discussions with
J.L. Gervais and G. Weigt. H.N. would also like to thank J. Schnittger
for explaining his as yet unpublished results. \\
Y.K. is grateful to the generous support and the hospitality of the
members of the theory group of the Universit\"at Hamburg and DESY,
where a part of this work was performed.
  The research of Y.K. is supported in part
by the grant in aid from the Japanese Ministry of Education,
No.04245208 and No.04640283.
%%%%%%%%%%%%%%%%%%%%%%%%%%%%%%%%%%%%%%%%%%%%%%%%
%%%%%%%%%%%%%%%%%%%%%%%%%%%%%%%%%%%
% lvappend.tex
%%%%%%%%%%%%%%%%%%%%%%%%%%%%%%%%%%%
\setcounter{equation}{0}
\renewcommand{\theequation}{A.\arabic{equation}}
\vspace{1.2cm}\parn
{\Large\bf Appendix} \qquad
{\bf Euclidean-Minkowski Conversion} \parbigskipn
In this appendix, we summarize how  Euclidean and
Minkowski formulations of conformal field theories are related.  When
there is a background charge,  transcription is somewhat non-trivial.
We shall mostly deal  with the chiral sector.
\par
%%%%%%%%%%%%%%%%%%%%%%%%%%%%%%%%%%%%%%%%%%%%%%
We begin with the Euclidean case. With $z$ the \lq\lq plane" coordinate,
 the energy-momentum tensor is expanded as
\eqabegin
 T(z) &=& \sum_n L_n z^{-n-2} \comma
\eqaend
where $L_n$'s are assumed to satisfy the \lq\lq standard form" of the
Virasoro algebra
\eqabegin
 \left[L_m, L_n\right] &=& (m-n)L_{m+n} + {c\over 12}(m^3-m)
 \delta_{m+n,0} \period \label{stdVir}
\eqaend
A primary field $\phi(z)$ of dimension $\Delta$ is characterized by
\eqabegin
 \phi(z) &=& \sum_n \phi_n z^{-n-\Delta}\comma \\
 \left[ L_n, \phi(z)\right] &=&
  z^n\left(z{d\over dz}+(n+1)\Delta \right)\phi(z) \period
\eqaend
The \lq\lq cylinder" coordinates $\tilde{z},\,\bar{\tilde z}$,
 which will be directly related the Minkowski light-cone coordinates,
 are defined by the conformal tranformation
\eqabegin
 z &=& \re^{\tilde z}, \qquad \tilde z = \tau + i\sigma \comma
 \label{cylinder}\\
 \bar{z} &=& \re^{\bar{\tilde z}}, \qquad \bar{\tilde z}
   = \tau -i\sigma \period
\eqaend
Then the energy-momentum tensor is transformed into
\eqabegin
 T(\tilde z) &=& \left({dz \over {d\tilde z}}\right)^2 T(z)
    + {c\over 12}\left\{ z, \tilde z \right\} \comma \label{ctofT}
\eqaend
where $\left\{ z, \tilde z \right\}$ is the Schwarzian derivative
\eqabegin
 \left\{ z, \tilde z \right\} &=& {z'''\over z'} -{3\over 2}
  \left( {z'' \over z'}\right)^2 \\
 && ( z' = dz / { d\tilde z}, \qquad etc.)\period \nn
\eqaend
For the transformation (\ref{cylinder}) above, the Schwarzian
derivative is $-1/2$.  Thus we get
\eqabegin
 T(\tilde z) &=& z^2T(z) -{c\over 24}\comma \\
  L^{\tilde z_n} &=& L_n -{c \over 24} \delta_{n,0} \period
\eqaend
The generators $L^{\tilde z}_n$ satisfy the algebra
\eqabegin
 \left[ L^{\tilde z}_m, L^{\tilde z}_n\right]
   &=& (m-n)L^{\tilde z}_{m+n}+ {c\over 12}m^3\delta_{m+n,0} \nn
\eqaend
that is, {\it without} a central term linear in $m$. \par
The primary field $\phi(z)$ is transformed into
\eqabegin
 \phi(\tilde z) &=& \left({dz\over {d\tilde z}}\right)^\Delta
  \phi(z) = \sum \phi_n z^n = \sum \phi_n \re^{-n\tilde z} \period
\eqaend
A simple calculation leads to
\eqabegin
 \left[ L^{\tilde z}_n, \phi(\tilde z)\right]
    &=& \re^{n\tilde z}\left( {d\over {d\tilde z}}+n\Delta\right)
      \phi(\tilde z) \period
\eqaend
Note that the factor in front of $\Delta$ is changed from $n+1$ to
$n$.  \par
%%%%%%%%%%%%%%%%%%%%%%%%%%%%%%%
Minkowski formulation is obtained from Euclidean cylinder formulation
by the replacement $\tau \longrightarrow i\tau$, which converts the
cylinder coordinates into the light-cone coordinates:
\eqabegin
 \tilde z = \tau_E + i\sigma\quad  &\longrightarrow &
       \quad i(\tau_M + \sigma) = i\xplus \comma\\
 \bar{\tilde z} = \tau_E - i\sigma\quad  &\longrightarrow &
       \quad i(\tau_M - \sigma) = i\xminus \comma \\
 iz\del & \longrightarrow & \quad \delplus \equiv {\del \over
     \del\xplus} \comma \\
 i\bar{z}\bar{\del}  & \longrightarrow & \quad \delminus
  \equiv {\del \over \del\xminus} \period
\eqaend
( Here and hereafter,  $\del$ ($\bar{\del}$) means $\del/\del z$
($\del/\del \bar{z}$).)  Using these formulae, it is easy to obtain
 operator product expansions, such as (\ref{TT}), from the corresponding
 ones in the Euclidean plane coordinate formulation.  \par
%%%%%%%%%%%%%%%%%%%%%%%%%%%%%%%%%%%%
Before discussing the important case of a free boson with a background
charge, let us briefly describe the notion of marginal deformation of
Virasoro generators.
Let $L_m$'s satisfy the standard form of the
Virasoro algebra (\ref{stdVir}).
 We also suppose that there exists a $U(1)$ type current with the mode
 oscillators $\alpha_m$, which satisfy
\eqabegin
 \left[ \alpha_m, \alpha_n \right] &=& m\delta_{m+n,0} \comma \\
 \left[ L_m, \alpha_n\right] &=& -n\alpha_{m+n}
   +(A_1m + A_2 m^2) \delta_{m+n,0} \period
\eqaend
Note that we allow for a term proportional to $\mndelta$ in the
second equation.  This will occur later when we deal with a system
with a background charge. \par
With this setting one can deform the Virasoro generator in the following
 manner:
\eqabegin
 \Ltilde_m &\equiv & L_m + (B_0 +B_1m) \alpha_m + C \delta_{m,0} \period
\eqaend
Then a straightforward calculation shows
\eqabegin
 \left[ \Ltilde_m, \Ltilde_n\right] &=& (m-n)\Ltilde_{m+n}
   +{c\over 12}(m^3-m) \delta_{m+n,0} \nn\\
 & & \quad - m^3B_1(2A_2 +B_1)\mndelta
   + m(B_0^2+2A_1B_0-2C) \delta_{m+n,0} \period
\eqaend
Thus if we take
\eqabegin
 A_2 &=& \bracetwocases{-{B_1 \over 2}}{\qquad \mbox{if $B_1 \ne 0$ }}
{\mbox{arbitrary}}{\qquad \mbox{if $B_1 =0$ }} \\
 C &=& {B_0^2 \over 2} + A_1B_0 \comma
\eqaend
$\Ltilde_m$ satisfy the standard form of the Virasoro algebra with
the same central charge. \parmedskipn
%%%%%%%%%%%%%%%%%%%%%%%%%%%%%%%%%%%%%%%%%%%%%%%%%%%%%%%%%%%
Now let us  consider a system of a free boson $\phi$ with a background
charge $q$.
In Euclidean plane coordinate formulation, $\phi(z,\bar{z})$ is
expanded as
\eqabegin
 \phi(z,\bar{z}) &=& \phi_0 -i\alpha_0 \ln(z\bar{z})
   + \overi \sum_{n\ne 0}{1\over n}\left( \alpha_{-n}z^n +
\bar{\alpha}_{-n}\bar{z}^n \right) \comma
\eqaend
where the mode operators satisfy the usual commutation relations.
 The holomorphic energy-momentum tensor is given by
\eqabegin
 T(z) &=& -\half (\del \phi)^2 +q\del^2\phi
   \equiv \sum L^E_n z^{-n-2} \comma\\
 L^E_n &=& L^{(0)}_n + iq(n+1)\alpha_n \comma \label{LE} \\
 L^{(0)} &=& \half \sum :\alpha_{n-m}\alpha_m : \period
\eqaend
$L^E_n$ satisfy the standard form of the Virasoro algebra with
 $c =1+12 q^2 $. \par
It is important to note that  $\phi$ transforms
under conformal transformation with an additional inhomogeneous term
 proportional to the background charge $q$:
\eqabegin
 \left[L^E_n, \phi(z)\right] &=& z^n\left( z\del \phi(z) + q(n+1)\right)
\period \label{ctofphi}
\eqaend

Now we make a  conversion to the Minkowski formulation.  Due to the
the non-trivial tranformation property just mentioned, $\phi$ undergoes
the replacement
\eqabegin
 \phi(z,\bar{z}) &\longrightarrow& \phi(\xplus,\xminus)
-iq(\xplus+\xminus) -i\pi q \period
\eqaend
In terms of modes, this is equivalent to the shifts
\eqabegin
 \phi_0 &\longrightarrow & \phi_0^M -iq\pi \comma\nn\\
 \alpha_0 &\longrightarrow & \alpha_0 -iq \comma\label{phishift}\\
 \alpha_n &\longrightarrow& \alpha_n \period   \nn
\eqaend
To convert the energy-momentum tensor, one must take into account this
replacement in addition to the transformation (\ref{ctofT}). After
a simple calculation we obtain
\eqabegin
 T(\xplus) &=& \half (\delplus\phi)^2 -q\delplus^2\phi   \\
  &\equiv& \sum L^M_n \re^{-in\xplus} \comma \\
 L^M_n &=& L^{(0)} + iqn\alpha_n = L^E_n -iq\alpha_n \comma \label{LM}
\eqaend
where the inessential overall additive constant $-1/24$ has been
dropped.
$L^M_n$ satisfies the Virasoro algebra of the form
\eqabegin
 \left[ L^M_m, L^M_n \right]
  &=& (m-n)L^M_{m+n} +\delta_{m+n,0}
   \left( {c\over 12}(m^3-m) + q^2 m \right) \comma
\eqaend
which is not of the standard form.  But since the difference is linear
 in $m$, we can shift $L^M_0$ to cancel this term.  It is easy to see
 that
\eqabegin
 \tilde{L}^M_n &=& L^M_n + {q^2\over 2}\delta_{n,0} \\
   &=& L^E_n -iq\alpha_n + {q^2\over 2}\delta_{n,0}
\eqaend
satisfies the standard form. Notice that
\eqabegin
 \left[ L^E_m, \alpha_n \right] &=& -n\alpha_{m+n}
     + iqm(m+1)\delta_{mn} \period
\eqaend
This is of the form treated in the discussion of marginal deformation
 of the Virasoro algebra.  Indeed we can identify
\eqabegin
 A_1 &=& iq, \qquad A_2 = iq \comma\nn\\
 B_0 &=& -iq, \qquad B_1 = 0 \comma\\
 C &=& {q^2\over 2} \comma\nn
\eqaend
and  easily check that these coefficients precisely satisfy the
conditions for the marginal deformation. \par
It is extremely useful to regard the shift (\ref{phishift}) as
a similarity tranformation $\calU$ defined by
\eqabegin
 \calU &=& \re^{-q\phi_0}\re^{q\pi\alpha_0} \period
\eqaend
Then the conversion of $\phi$ can be  succinctly expressed as
\eqabegin
 \phi &\longrightarrow& \calU\, \phi \,\calUinv \period
\eqaend
Moreover, it is not difficult to check explicitly that this operation
correctly converts the Virasoro generators.  Namely,
\eqabegin
\calU \, L^E_n \, \calUinv &=&  L^M_n \comma
\eqaend
where $L^E_n$ and $L^M_n$ are as defined in (\ref{LE}) and
(\ref{LM}). This makes obvious the previously mentioned fact
that both $L^E_n$ and $L^M_n$ satisfy the standard form of the
Virasoro algebra. \par
As an application of the similarity transformation, let us give the
conversion formula for the exponential operator $\re^{\lambda\phi}$
 carrying the conformal weight $\Delta = -(\lambda^2/2)+\lambda q$.
If we denote by $:\ \ :_E$ and $:\ \ :_S$ the Euclidean and the
symmetric normal orderings respectively,  a simple
calculation gives
\eqabegin
\calU  :\re^{\lambda\phi(z,\bar{z})}:_E
\left({dz \over d\xplus}{d\bar{z} \over d\xminus}\right)^\Delta
  \calUinv &=& :\re^{\lambda\phi}:_S \re^{-i\pi \lambda^2 /2} \period
\eqaend
Apart from a coordinate independent phase factor, the usual Euclidean
 normal ordering is precisely converted to the symmetric normal ordering
  This clearly shows the necessity of symmetric normal ordering adopted
 in the text.
%%%%%%%%%%%%%%%%%%%%%%%%%%%%%%%%%%%%%%%%%%%%%
%%%%%%%%%%%%%%%%%%%%%%%%%%%
% lvref.tex
%%%%%%%%%%%%%%%%%%%%%%%%%%%

%%%%%%%%%%%%%%%%%%%%%%%%%%%%%%%%%%%%%%%%%%%%%%%%
\end{document}